\documentclass[draft]{agujournal2019}
\usepackage{url} 
\usepackage{lineno}
\usepackage[inline]{trackchanges} 
\usepackage{soul}
\usepackage[flushleft]{threeparttable}
\usepackage{ragged2e}
\linenumbers
\draftfalse
\journalname{Journal of Advances in Modeling Earth Systems (JAMES)}

\begin{document}
\nolinenumbers
\justifying

\title{Critical role of vertical radiative cooling contrast in triggering episodic deluges in small-domain hothouse climates}

%
%




\authors{Xinyi Song\affil{1}, Dorian S. Abbot\affil{2}, Jun Yang\affil{1}}

 \affiliation{1}{Laboratory for Climate and Ocean-Atmosphere Studies, Department of Atmospheric and Oceanic Sciences, School of Physics, Peking University, Beijing, China}
 
 \affiliation{2}{Department of the Geophysical Sciences, The University of Chicago, Chicago, USA}





\correspondingauthor{Jun Yang}{junyang@pku.edu.cn}


\begin{keypoints}
\item Lower-tropospheric radiative heating is unnecessary for the occurrence of episodic deluges
\item The strong vertical gradient of radiative cooling is a key factor in triggering episodic deluges
\item The occurrence of episodic deluges needs strong convective inhibition (CIN)


\end{keypoints}


\begin{abstract}
\citeA{Seeley2021} showed that in small-domain cloud-resolving simulations the pattern of precipitation transforms in extremely hot climates ($\ge$ 320 K) from quasi-steady to organized episodic deluges, with outbursts of heavy rain alternating with several dry days. They proposed a mechanism for this transition involving increased water vapor absorption of solar radiation leading to net lower-tropospheric radiative heating. This heating inhibits lower-tropospheric convection and decouples the boundary layer from the upper troposphere during the dry phase, allowing lower-tropospheric moist static energy to build until it discharges, resulting in a deluge. We perform cloud-resolving simulations in polar night and show that the same transition occurs, implying that some revision of their mechanism is necessary. We show that episodic deluges can occur even if the lower-tropospheric radiative heating rate is negative, as long as the magnitude of the upper-tropospheric radiative cooling is about twice as large.  We find that in the episodic deluge regime the mean precipitation can be inferred from the atmospheric column energy budget and the period can be predicted from the time for radiation and reevaporation to cool the lower atmosphere.

\end{abstract}

\section*{Plain Language Summary}
Precipitation plays an important role in Earth's climate and habitability, and also influences important weathering processes such as carbonate-silicate cycle. In the distant future, Earth may experience a very hot and wet ``hothouse" climate, just like that in the past Archean. Modelling results show that in a hothouse climate, precipitation transforms into an ``episodic deluge" pattern, with outbursts of heavy rain alternating with several dry days. In this study, we find that positive lower-tropospheric heating is not the necessary cause for episodic deluges. Instead, vertical radiative cooling contrast is critical in triggering the episodic deluges in small-domain hothouse climates. We also try to understand the underlying mechanism of episodic deluges through CIN and CAPE analyses. 

\section{Introduction} \label{sec:intro}
Earth might have experienced an extremely warm and wet climate, a ``hothouse," in the Archean \cite{Sleep_2010,Charnay_2017,steffen2018trajectories}, or in the aftermath of a snowball Earth event \cite{Higgins_2003,Le_Hir_2009,Pierrehumbert_2011,hoffman2017snowball,yang2017persistence}, and may experience a hothouse climate again in the distant future \cite{Ingersoll_1969,kasting1984response,Kidder_2012,Goldblatt_2013,Leconte_2013,Ramirez_2014}. Previous work mainly used general circulation models (GCMs), and concluded that there could be a lower-tropospheric temperature inversion and significant increase in upper-tropospheric cloud cover in hothouse climates \cite{Wordsworth_2013,Wolf_2015,Popp_2016,wolf2018evaluating}. \citeA{Seeley2021} moved beyond GCMs by using convective scale cloud-resolving models and found that the precipitation would organize in time into an ``episodic deluge" pattern. In this regime, the majority of the grid points have an outburst of heavy rain at the same time (Figure 4d in \citeA{Seeley2021}), followed by several dry days. \citeA{Seeley2021} investigated this in three different cloud-resolving models, Das Atmosphärische Modell (DAM) (Romps, 2008), the System for Atmospheric Modeling (SAM) \cite{Khairoutdinov_2003}, and the Cloud Model 1 (CM1) (Bryan and Fritsch, 2002), and modified the radiative transfer scheme of DAM in order to be more accurate in hot climates. They verified that the onset of episodic deluges does not depend on the specific model choice. Most of their experiments are in a small domain of 72 km $\times$ 72 km, and the episodic deluge is a synchronized behavior between the grid points.


\begin{figure}[!ht]
\noindent
    \includegraphics[width=1.0\textwidth]{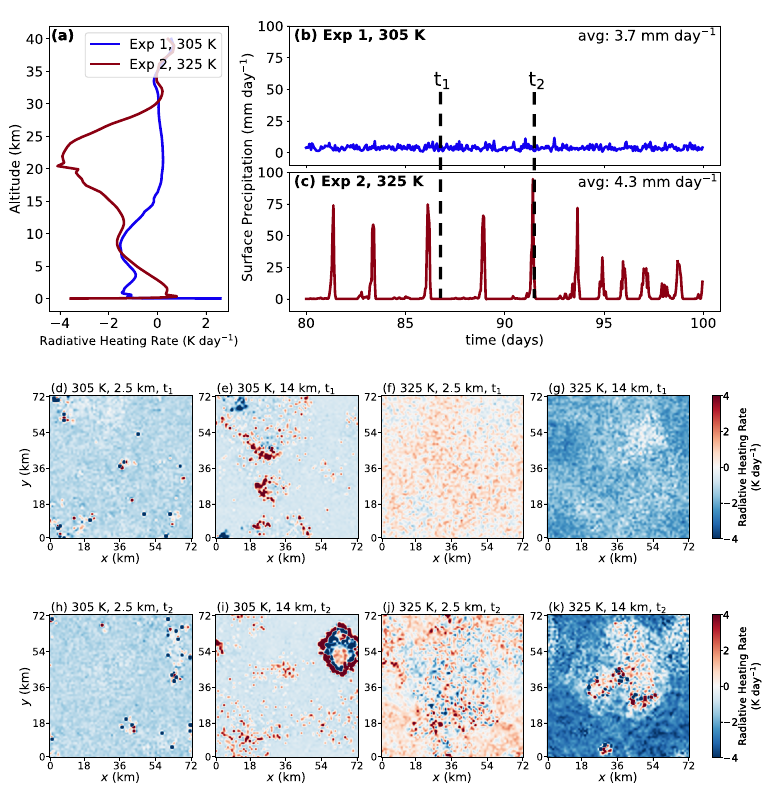}
    \caption{Reproduction of modelling results in \citeA{Seeley2021} using SAM. Panel (a) shows time-averaged radiative heating rate. Panels (b) and (c) show the precipitation pattern in two simulations with the surface temperature fixed at 305 K (Exp 1) and 325 K (Exp 2), respectively. Panels (d)--(g) show the horizontal distribution of the radiative heating rate (longwave plus shortwave) for both cases during a dry spell (t$_1$). Panels (h)--(k) show the horizontal distributions during a convection (t$_2$). Each case shows slices at 2.5 km and 14 km. For time variations, see the video version in Supporting Information. Both cases run for 100 days.}
    \label{fig_reproduction}
\end{figure}

\citeA{Seeley2021} argued that episodic deluges are mainly caused by lower-tropospheric radiative heating: When the climate is warm enough, more lower-tropospheric water vapor increases shortwave absorption, and results in a net positive heating rate. Figure~\ref{fig_reproduction} shows our reproduction of the results of \citeA{Seeley2021} using the cloud-resolving model SAM. When the surface temperature is 305 K (Exp 1), there is net radiative cooling in the lower troposphere (blue line in Figure~\ref{fig_reproduction}a), and the precipitation pattern is quasi-steady (Figure~\ref{fig_reproduction}b). When the surface temperature is 325 K (Exp 2), the lower-tropospheric radiative heating rate becomes positive (red line in Figure~\ref{fig_reproduction}a), and episodic deluges occur. The full mechanism proposed by \citeA{Seeley2021} to explain episodic deluges, starting in a dry phase, is as follows:  lower-tropospheric radiative heating inhibits convection in the lower troposphere. Strong radiative cooling in the upper troposphere leads to condensation and elevated precipitation above an ``inhibition layer". Most droplets of upper-tropospheric precipitation reevaporate in the relatively warm inhibition layer, so very little (or zero) precipitation makes it to the surface. As time goes by, reevaporation of precipitation cool down the inhibition layer to the point that inhibition is broken, triggering strong convection. After the convection, inhibition starts again, and the cycle continues. 


The heating rate profiles undergo three major changes as the surface temperature increases to 325 K. First, the lower troposphere shifts from cooling to heating. Second, the vertical gradient of the heating rate profile increases (Figure \ref{fig_reproduction}a, see also Figure 2b in \citeA{Seeley2021}). When the surface temperature is 305 K, the radiative heating rate is about $-$1.5 K day$^{-1}$ from the near surface layer to about 10 km, then smoothly transits to 0 K day$^{-1}$ in the stratosphere (blue line in Figure~\ref{fig_reproduction}a). When the surface temperature is 325 K, the lower-tropospheric radiative heating rate is about 0.5 K day$^{-1}$ at about 2 km, but the upper-tropospheric radiative heating rate is about $-$4 K day$^{-1}$ at 20 km (the red line in Figure~\ref{fig_reproduction}a). The increased surface temperature leads to a warmer upper troposphere and higher water vapor concentration. Both factors intensify longwave cooling in the upper troposphere. Third, during the dry period, the horizontal distribution of radiative heating  is much more homogenous for the 325 K case, both at the lower troposphere (Figure \ref{fig_reproduction}f) and in the upper troposphere (Figure \ref{fig_reproduction}g). Note that the heating rates for both cases are fairly heterogeneous horizontally during convection (Figure~\ref{fig_reproduction}h--k). For the video version of Figure~\ref{fig_reproduction}, see Supporting Information Video S1.

\citeA{Seeley2021} pointed to lower-tropospheric radiative heating as the primary factor leading to episodic deluges. In this paper, we suggest instead that the vertical gradient of radiative cooling is a more important factor for the onset of episodic deluges. This point is emphasized by the fact that episodic deluges occur even when the radiative heating rate in the lower troposphere is negative (Section~\ref{sec:polarnight}). More specifically, episodic deluges require that the magnitude of radiative cooling in the upper-troposphere be about twice that in the lower troposphere (Section~\ref{subsec:mechanism}). The episodic deluge regime the mean precipitation can be inferred from the atmospheric column energy budget, and the period can be predicted from the time for radiation and reevaporation to cool the lower troposphere.




\section{Episodic deluges during polar night} \label{sec:polarnight}

All the experiments in this study use version 6.11.6 of SAM \cite{Khairoutdinov_2003,Khairoutdinov_2018}, one of the three models used in \citeA{Seeley2021}. The horizontal resolution in each experiment is 2 km with 72 grid points in each direction. The vertical resolution is 144 grid points within 64 km. CO$_2$ is set to 400 ppmv and the experiments do not contain ozone. The time step is 10 seconds, and the output statistics are hourly averages.

\begin{figure}[!ht]
\noindent
    \includegraphics[width=1.0\textwidth]{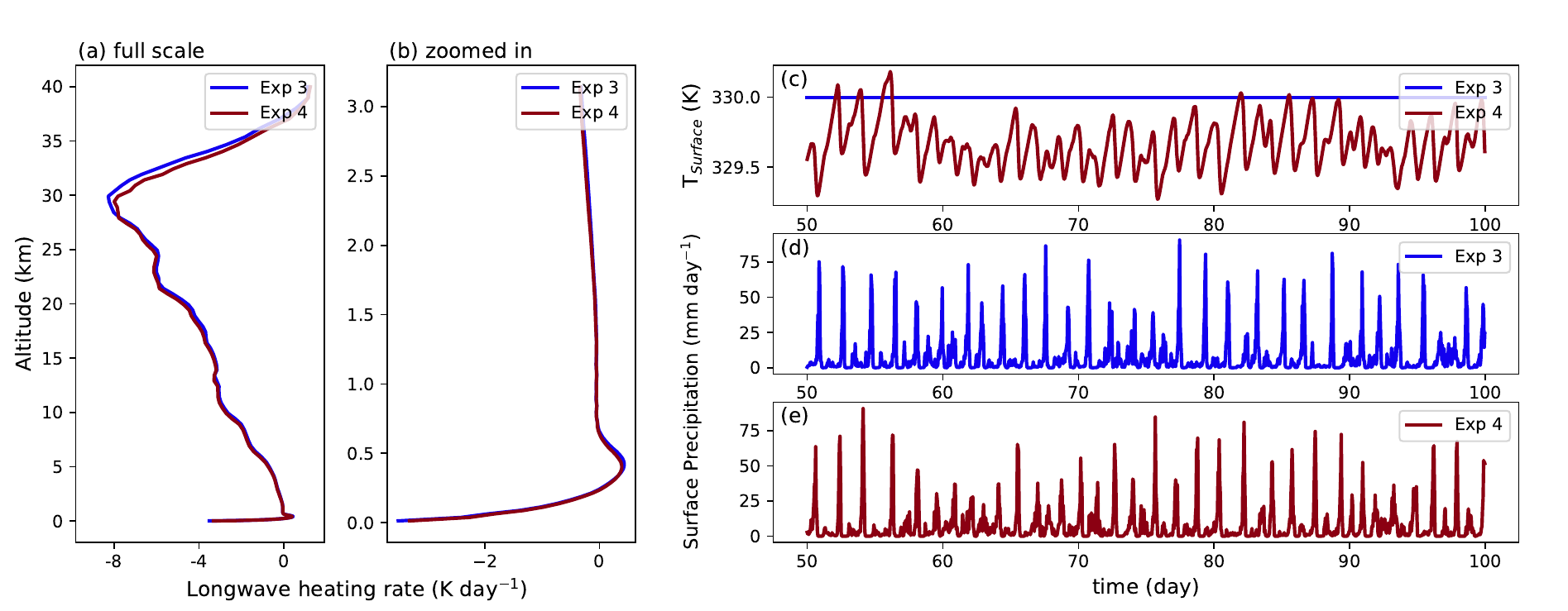}
    \caption{Full scale (a) and zoomed in (b) radiative heating rate profiles in polar night hothouse climate. One experiment fixes the sea surface temperature (SST) at 330 K (Exp 3). The other includes an ocean heat import of 230 W m$^{-2}$ (Exp 4), in order to maintain the SST at around 330 K (c). Episodic deluges occur in both experiments (d and e).}
    \label{fig_polarnight}
\end{figure}

We find episodic deluges in polar night simulations, demonstrating that shortwave heating is not necessary for episodic deluges (Figure~\ref{fig_polarnight}). We find polar night episodic deluges both with a fixed sea surface temperature of 330~K (Exp 3) and with a two-meter-deep slab ocean and a positive ocean heat flux of 230 W m$^{-2}$ (Exp 4), which produces a similar sea surface temperature. In both cases there is a deluge, or convective, period that lasts several hours and has a peak precipitation of about 80 mm day$^{-1}$, ten times the average rate. This is followed by a dry, inhibition period that lasts several days during which the precipitation rate is usually below 2 mm day$^{-1}$.

\begin{figure}[!ht]
\noindent
    \includegraphics[width=1.0\textwidth]{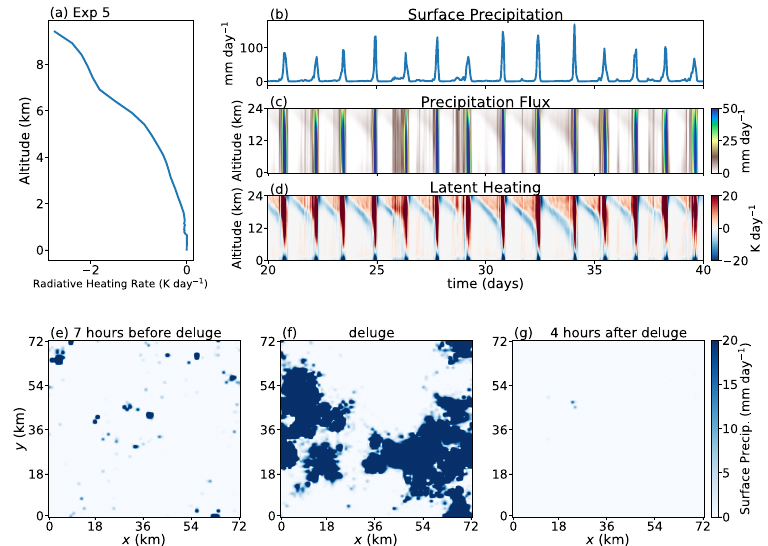}
    \caption{Prescribed radiative heating rate with lower-tropospheric radiative heating removed (a) and simulated precipitation (b) during polar night. Panels (c) and (d) show the altitude-time plot of precipitation flux and latent heating. Panels (e)–(g) show snapshots of surface precipitation at 7 hours before a deluge, during the deluge, and 4 hours after the deluge, respectively. Episodic deluges still exist even if there is no lower-tropospheric radiative heating.}
    \label{fig_noheating}
\end{figure}



So far, the polar night simulations indicate that shortwave heating is not a necessary condition for episodic deluges, but we still cannot rule out the necessity of lower-tropospheric radiative heating, as the heating rate at about 0.5 km is slightly positive (Figures ~\ref{fig_polarnight}a \& b). Random process, transient temperature inversion, or some unknown processes may cause this longwave heating. To exclude the influence of this longwave heating, we add an additional experiment, Exp 5, to smooth the near surface layer of the heating rate  profile in the polar night experiment, and set the maximum value to 0 K day$^{-1}$, as shown in Figure~\ref{fig_noheating}a. Episodic deluges still occur under this radiative heating profile (Figure~\ref{fig_noheating}b). Therefore, we can confirm that lower-tropospheric radiative heating is not necessary for episodic deluges.

What is necessary to cause episodic deluges? Let's take a look back to Figure \ref{fig_reproduction}a. Another important feature that changes between the 305 K and 325 K cases 
is the vertical gradient in the heating rate: the upper-tropospheric cooling is much stronger in the 325 K case, leading to a larger vertical gradient. In the following sections, we use modelling experiments to investigate how different heating rate profiles can influence the episodic deluges. 


\section{Heating Rate Profile Experiments and Results}
\begin{figure}[ht]
\noindent
    \centering\includegraphics[width=0.5\textwidth]{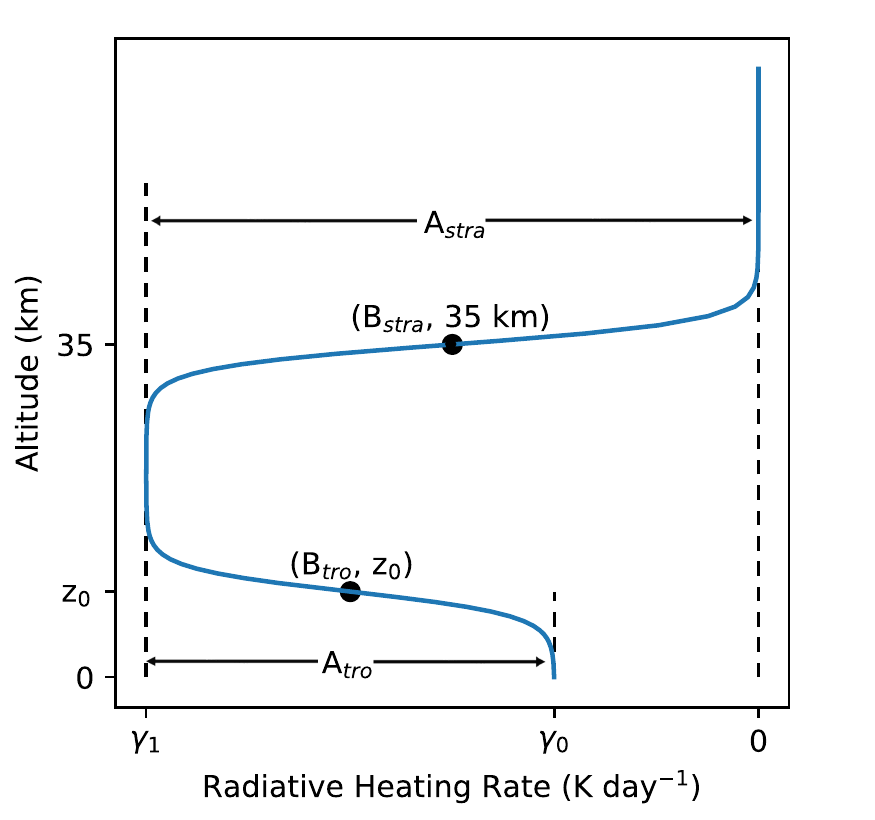}
    \caption{Illustration of the prescribed radiative heating rate profile. The profile can be divided into three parts. The lower troposphere spans from surface to $z_0$, with radiative heating rate of $\gamma_0$. The upper troposphere, with radiative heating rate of $\gamma_1$, spans from $z_0$ to 35 km. The stratosphere spans from 35 km to the model top with radiative heating rate of 0 K day$^{-1}$. The layers are connected to each other smoothly.}
    \label{fig_profile}
\end{figure}

The following experiments are run for 100 days with a fixed sea surface temperature of 325 K. All experiments restart from the 325 K reproduction simulation Exp 2. Considering the features of the radiative heating rate profile under 325 K, the profile can be roughly divided into three parts: the lower troposphere, the upper troposphere with strong radiative cooling, and the stratosphere. We prescribe the radiative heating rate profiles in a three-layer structure as shown in Figure~\ref{fig_profile}. The heating rate profiles do not evolve with time, so both longwave and shortwave radiative transfer are turned off in the model. The heating rate profiles are given by two functions, 

\begin{linenomath*}
\begin{equation}
  \gamma = \cases{
  A_{tro} \frac{a^{-z^*}-a^{z^*}} {a^{-z^*}+a^{z^*}} + B_{tro},  & for the troposphere, \cr
  A_{stra} \frac{a^{-z^*}-a^{z^*}} {a^{-z^*}+a^{z^*}} + B_{stra}, & for the stratosphere,
                }
    \label{RADQR}
\end{equation}
\end{linenomath*}
where $\gamma$ is the radiative heating rate in the units of K day$^{-1}$, $z^*$ is a relative height explained below, $a$ is a dimensionless number that controls the smoothness of the profiles, and here $a=1.5$. Coefficients $\gamma_0$, $\gamma_1$, and $z_0$ control the width and the central point of the heating rate profile, as shown in Figure \ref{fig_profile}. In the troposphere, $z^*=z-z_0$, $A_{tro} = {\gamma_0 - \gamma_1}$, and $B_{tro} = \frac{\gamma_0 + \gamma_1}{2}$. In the stratosphere, $z^*=z-35$, $A_{stra} = {\gamma_1}$, and $B_{stra} = \frac{0 + \gamma_1}{2}$. 

In the following sections we conduct 4 groups of experiments (Table~\ref{table_cin}) focusing on three factors: experiments in group one (G1) focus on the influence of the lower-tropospheric radiative heating rate $\gamma_0$; experiments in G2 and G3 focus on the influence of the upper-tropospheric radiative heating rate $\gamma_1$; experiments in G4 focus on the height of the inhibition layer $z_0$.


\subsection{The effect of lower-tropospheric heating rate ($\gamma_0$)} \label{subsec:x0}
\begin{figure}[ht]
\noindent
    \includegraphics[width=1.0\textwidth]{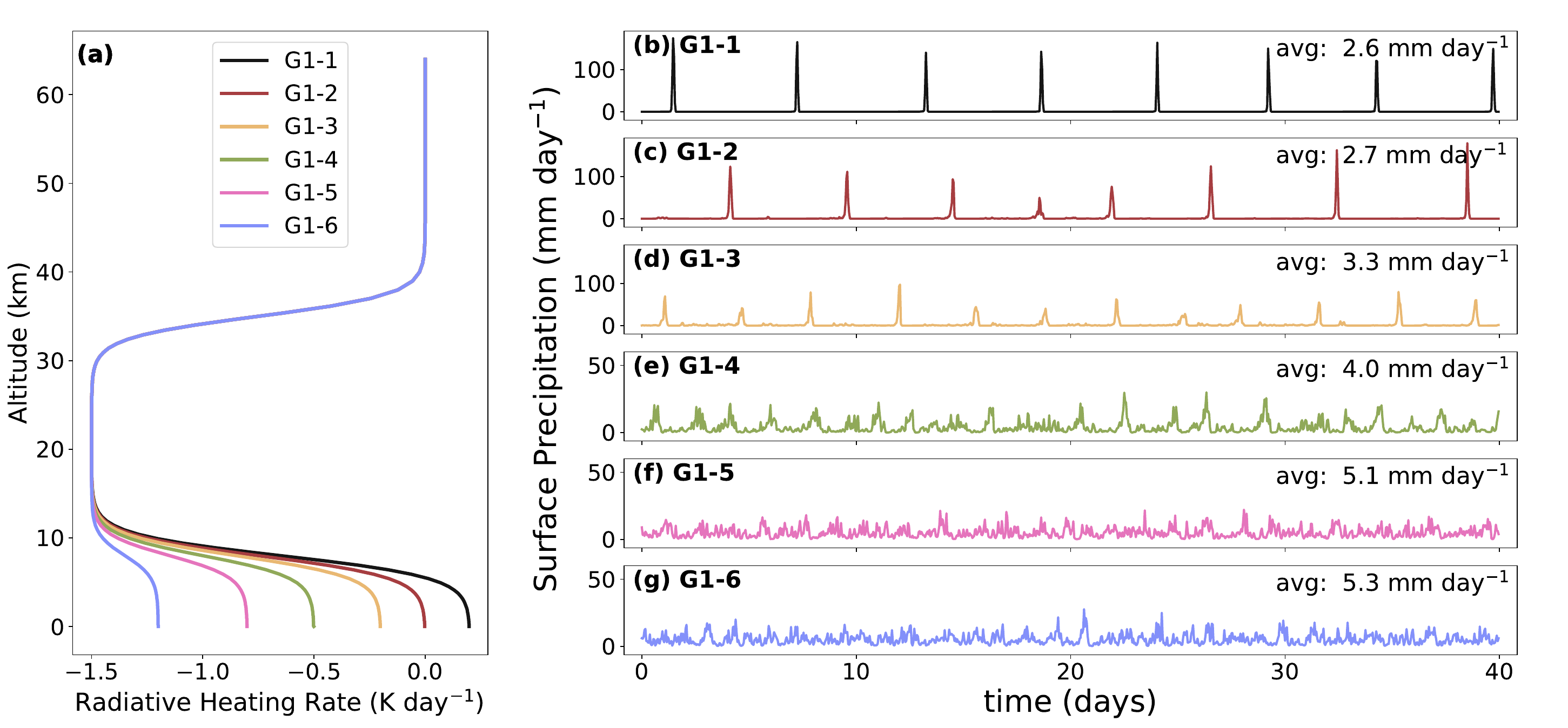}
    \caption{Simulations with fixed upper-tropospheric heating rate and different lower-tropospheric heating rates. Panel (a) shows the prescribed radiative heating rate profiles. The upper-tropospheric heating rate is $-$1.5 K day$^{-1}$. The lower-tropospheric heating rates in panels (b) to (g) are 0.2, 0, $-$0.2, $-$0.5,  $-$0.8, and $-$1.2 K day$^{-1}$, respectively. The inhibition layer height is 8 km.}
    \label{fig_lower_layers}
\end{figure}

First, we check the influence of the lower-tropospheric radiative heating rate while controlling other factors. Figure \ref{fig_lower_layers}a shows the radiative heating rate profiles in the first group of experiments (G1). The stratospheric heating rate is 0 K day$^{-1}$ and the upper-tropospheric heating rate $\gamma_1$ is $-$1.5 K day$^{-1}$. The inhibition layer height $z_0$ is 8 km. The lower-tropospheric heating rates $\gamma_0$ are 0.2, 0, $-$0.2, $-$0.5,  $-$0.8, and $-$1.2 K day$^{-1}$, respectively.

Figures \ref{fig_lower_layers}b--g show the surface precipitation. Episodic deluges happen even if the lower-tropospheric radiative heating rate is $-$0.2 K day$^{-1}$ (Figure \ref{fig_lower_layers}d), so again lower-tropospheric radiative heating is not required for episodic deluges. When the lower-tropospheric radiative heating rate is greater than $-$0.2 K day$^{-1}$, precipitation is strongly concentrated in the deluges  (Figures \ref{fig_lower_layers}b--d). Precipitation during the inhibition period is close to zero and the convective period is short. As the lower-tropospheric radiative heating rate slowly moves towards a negative value, the randomness of precipitation gradually increases into a quasi-steady pattern. When the lower-tropospheric radiative heating rate is less than $-$0.5 K day$^{-1}$, the precipitation pattern is no longer episodic but completely random (Figure \ref{fig_lower_layers}f \& g). 

We also carry out another set of experiments with heating rate profiles that are discontinuous at $z_0$ and at 35 km to exclude the influence of the transition details between the lower and upper troposphere. The heating rate profile settings are the same as those in G1, but the transitions are sharp (Figure \ref{fig_a1.x0}a). The results are similar (Figure \ref{fig_a1.x0}). The higher the lower-tropospheric radiative heating rate is, the more episodic the precipitation pattern is. Episodic deluges can occur even with a negative lower-tropospheric heating rate of $-$0.2 K day$^{-1}$. 


\subsection{The effect of upper-tropospheric heating rate ($\gamma_1$)} \label{subsec:x1}
\begin{figure}[ht]
\noindent
    \includegraphics[width=1.0\textwidth]{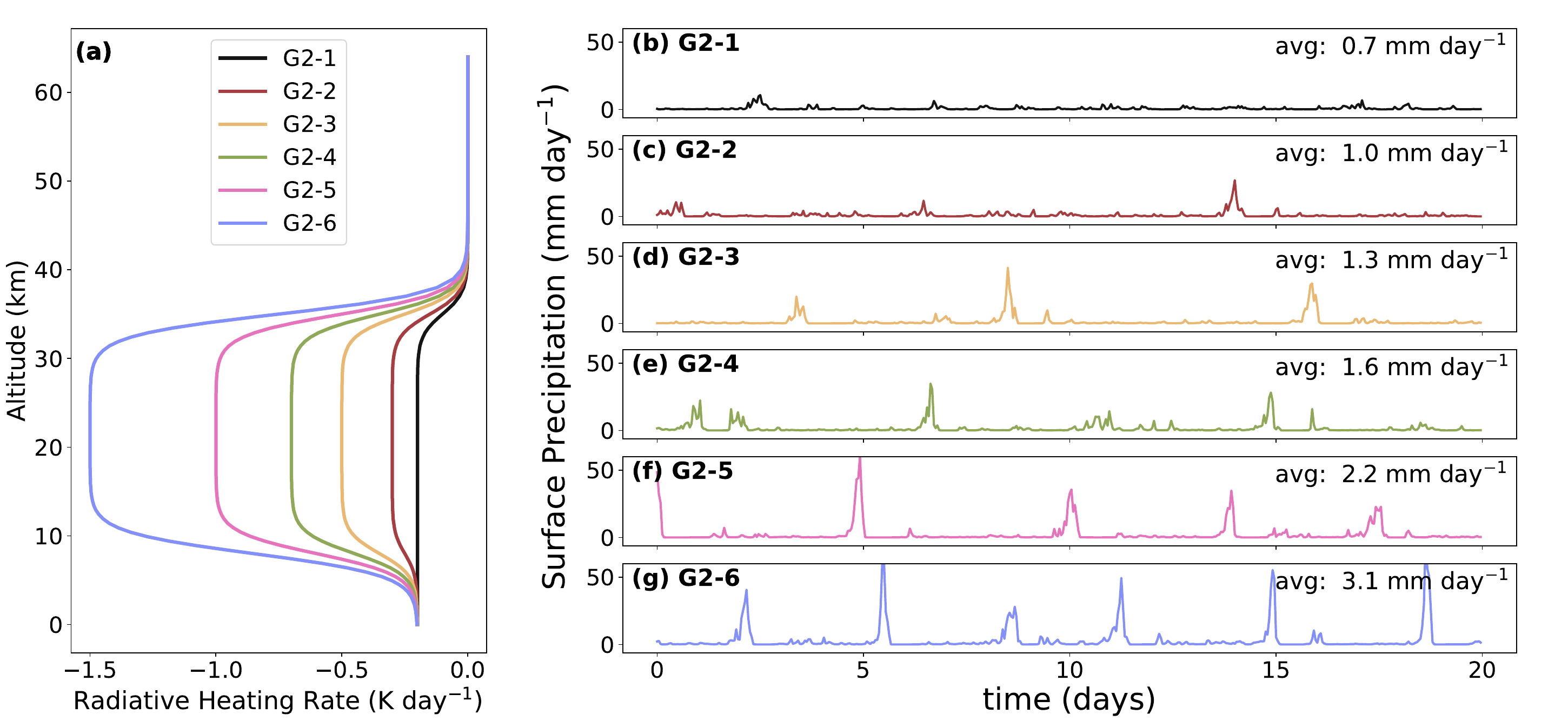}
    \caption{Simulations with fixed lower-tropospheric heating rate and different upper-tropospheric cooling rates. Panel (a) shows the prescribed radiative heating rate profiles. The lower-tropospheric heating rate is $-$0.2 K day$^{-1}$. The upper-tropospheric heating rates in panels (b) to (g) are $-$0.2, $-$0.3, $-$0.5, $-$0.7, $-$1.0, and $-$1.5 K day$^{-1}$, respectively. The inhibition layer height is 8 km.}
    \label{fig_middle_layers}
\end{figure}

Under what conditions will the negative lower-tropospheric radiative heating trigger episodic deluges? We examine the effect of upper-tropospheric heating rate while keeping the lower-tropospheric heating rate negative in the second group of experiments (G2). Figure \ref{fig_middle_layers}a shows the radiative heating rate profiles. The stratospheric heating rate is 0 K day$^{-1}$ and the lower-tropospheric heating rate, $\gamma_0$, is $-$0.2 K day$^{-1}$. The inhibition layer height $z_0$ is 8 km. The upper-tropospheric heating rates, $\gamma_1$, are $-$0.2, $-$0.3, $-$0.5, $-$0.7, $-$1.0, and $-$1.5 K day$^{-1}$, respectively. 

Figures \ref{fig_middle_layers}b--g show the surface precipitation. When the lower-tropospheric heating rate is negative, the upper-tropospheric cooling rate needs to be large enough to trigger episodic deluges. In this set of experiments, episodic deluges exist when the upper-tropospheric heating rate is less than $-$0.5 day$^{-1}$ (Figures \ref{fig_middle_layers}d--g). The stronger the upper-tropospheric cooling is, the shorter the episodic period is, and the more concentrated the precipitation is in the deluges. 

We also carry out a set of similar experiments with discontinuous heating rate profiles. The radiative heating rates for the lower troposphere, upper troposphere, and stratosphere are exactly the same as in G2, only the transitions between layers are sharp. The results are similar (Figure \ref{fig_a2.x1}) except for the exact period of episodic deluges. In general, periods with discontinuous heating rate profiles are longer than periods with smoothed profiles.

\begin{figure}[!ht]
\noindent
    \includegraphics[width=1.0\textwidth]{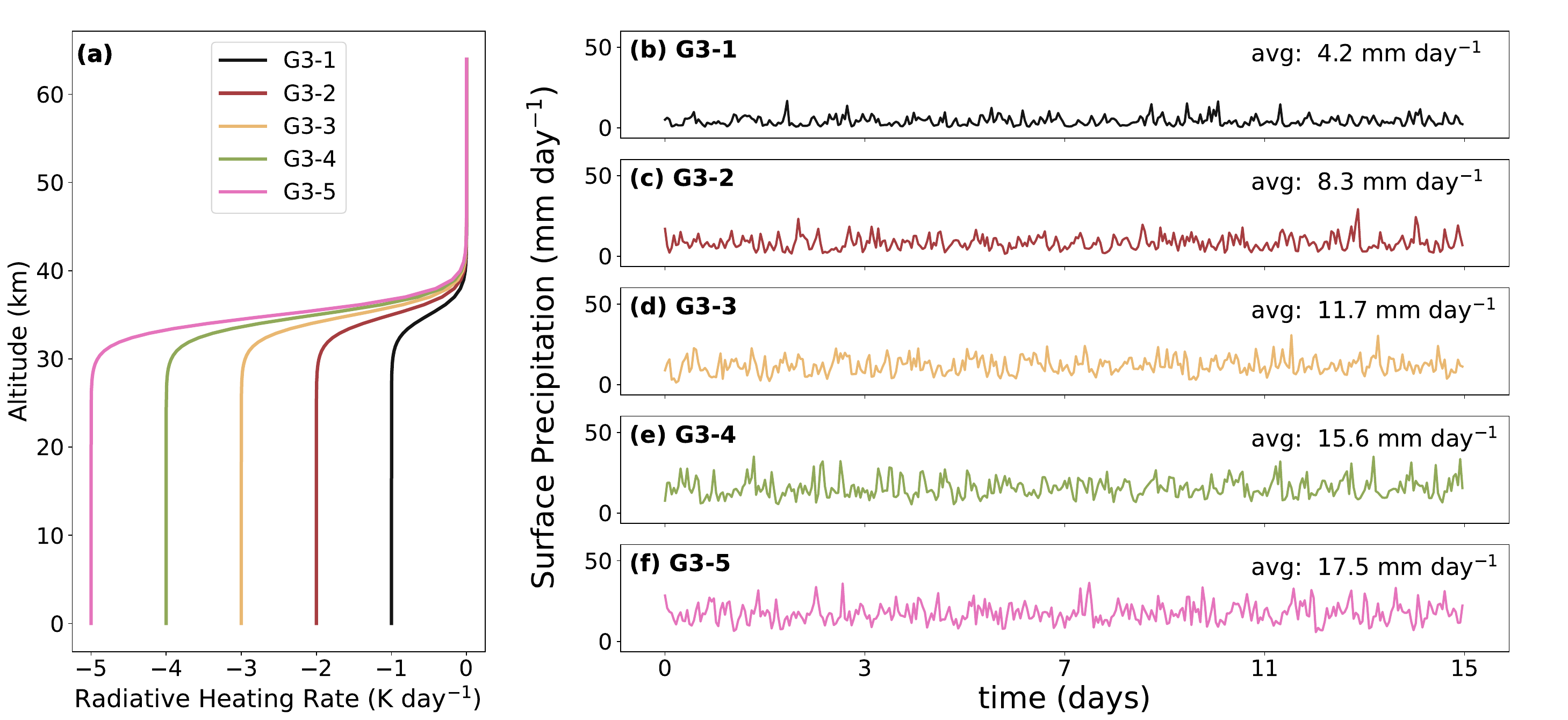}
    \caption{Simulations when lower-tropospheric heating rates and upper-tropospheric cooling rates are the same. Panel (a) shows the prescribed radiative heating rate profiles. The lower-tropospheric heating rates and the mupper-tropospheric heating rates in panels (b) to (f) are  $-$1, $-$2, $-$3, $-$4, and $-$5 K day$^{-1}$, respectively.}
    \label{fig_uni_tanh}
\end{figure}

Figure~\ref{fig_middle_layers} suggests that the upper-tropospheric radiative cooling needs to be strong in order to trigger episodic deluges. What if the lower-tropospheric and upper-tropospheric radiative cooling are both strong? Our third group of experiments (G3) set the lower troposphere and upper troposphere to have the same heating rate, that is, $\gamma_0=\gamma_1$. The upper layer heating rate is 0 K day$^{-1}$. The lower-tropospheric and upper-tropospheric heating rates are $-$1, $-$2, $-$3, $-$4, and $-$5 K day$^{-1}$, respectively. The heating rate profiles are similar to those in Figure 2a in \citeA{Seeley2021}, but they focused on changing the troposphere height and we focus on changing the tropospheric heating rate. Figure~\ref{fig_uni_tanh} shows the results. Precipitation patterns in G3 are all quasi-steady, showing that uniform strong radiative cooling is not enough to trigger episodic deluges. This highlights the fact that the vertical gradient in radiative cooling is more important for episodic deluges than the magnitude of radiative cooling. We also carry out a set of similar experiments with discontinuous heating rate profiles. The precipitation patterns are also quasi-steady (Figure~\ref{fig_uni_stair}).

\subsection{The effect of the inhibition layer height ($z_0$)} \label{subsec:z0}

\begin{figure}[!ht]
\noindent
    \includegraphics[width=1.0\textwidth]{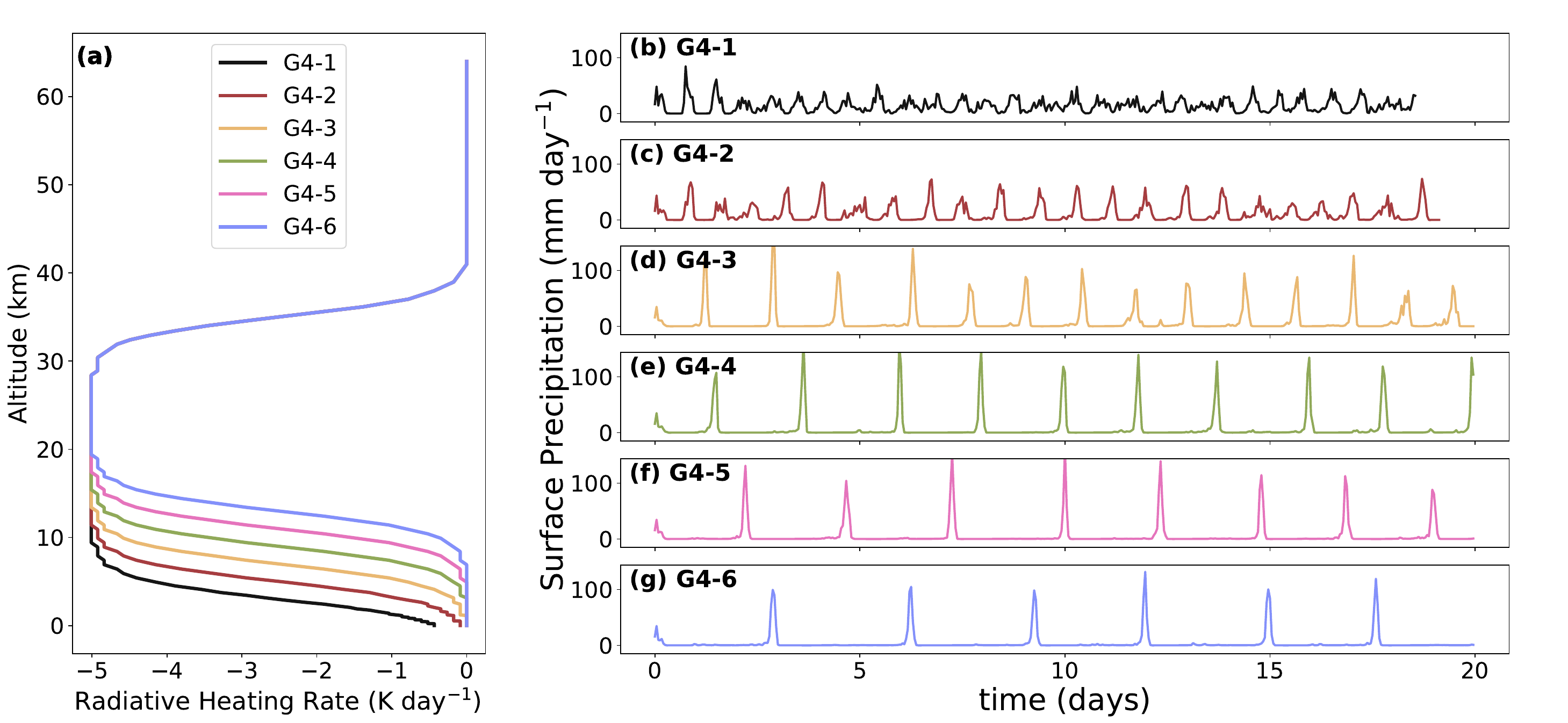}
    \caption{Simulations with fixed lower-tropospheric and upper-tropospheric heating rates but different inhibition layer heights. Panel (a) shows the prescribed radiative heating rate profiles. The lower-tropospheric heating rate is 0 K day$^{-1}$, and the upper-tropospheric heating rate is $-$5 K day$^{-1}$. $z_0$ in panels (b) to (g) are 3, 5, 7, 9, 11, and 13 km, respectively.}
    \label{fig_inhibition_height}
\end{figure}

The inhibition layer height is the last factor we test. Figure \ref{fig_inhibition_height}a shows the radiative heating rate profiles in the fourth group of experiments (G4). Both the upper and lower-tropospheric heating rates $\gamma_0$ are 0 K day$^{-1}$. The upper-tropospheric heating rate $\gamma_1$ is $-$5 K day$^{-1}$. The inhibition layer heights $z_0$ are 3, 5, 7, 9, 11, and 13 km, respectively.

Most of the cases have episodic deluges, but if the the inhibition layer is not high enough, for example, 3 km, precipitation is not very episodic (Figure \ref{fig_inhibition_height}b). As the inhibition height increases from 5 km to 13 km, the period increases from $\approx$ 1 day to $\approx$ 3 days. We will discuss the factors determining the period in section \ref{subsec:period}.

We also carry out a set of experiments with discontinuous heating rate profiles (Figure \ref{fig_a3.z0}) and the results are almost the same. The discontinuous heating rate profiles, compared with smoothed profiles, have somewhat higher inhibition heights (Figure \ref{fig_a3.z0}a vs Figure \ref{fig_inhibition_height}a). Because of higher inhibition heights, precipitation patterns with discontinuous heating rates (Figures \ref{fig_a3.z0}b \& c) are more episodic than the precipitation patterns with smoothed heating rates (Figures~\ref{fig_inhibition_height}b \& c).

\subsection{What is the underlying mechanism?}
\label{subsec:mechanism}
\begin{figure}[!htbp]
\noindent
    \centering
    \includegraphics[width=0.85\textwidth]{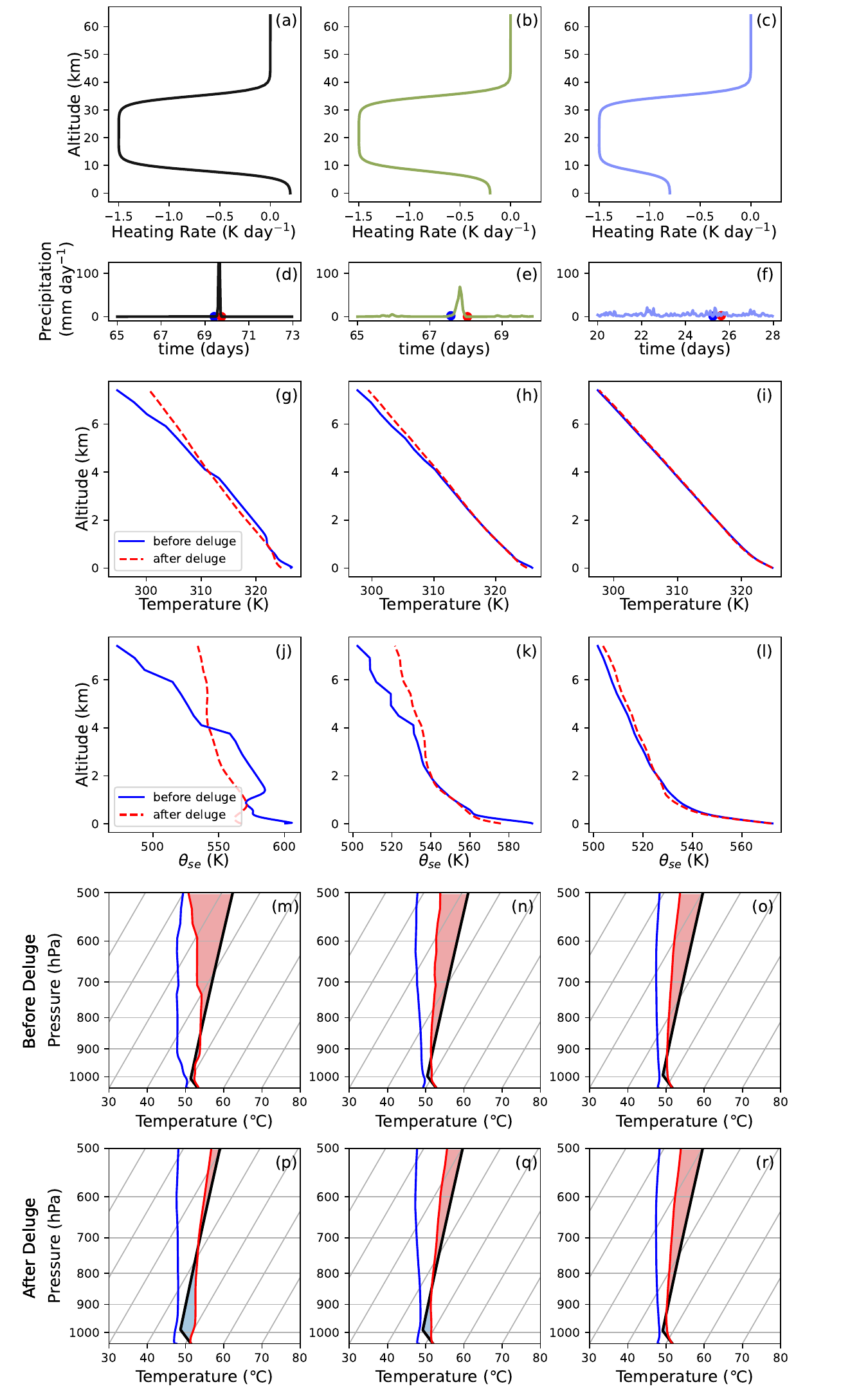}
    \caption{The specified radiative heating rate profiles, precipitation time series, air temperature, saturated equivalent potential temperature, and CAPE and CIN before and after convection. Panels in the left, middle, and right columns correspond to the experiments in Figures \ref{fig_lower_layers}b, \ref{fig_lower_layers}d, and \ref{fig_lower_layers}f, with lower-tropospheric heating rates of 0.2, $-$0.2, and $-$0.8 K day$^{-1}$, respectively. In the two bottom rows, the blue, black, and red lines are dew point temperature, parcel temperature, and environmental temperature, respectively. Red and blue shades in panels (m)-(o) show the CAPE and CIN before the deluge, and shades in panels (p)-(r) show the CAPE and CIN after the deluge.}
    \label{fig_mechanism}
\end{figure}

The results of the experiments above with prescribed heating rates can be divided into three types: episodic deluges with positive lower-tropospheric heating rate, episodic deluges with negative lower-tropospheric heating rate, and quasi-steady precipitation with negative lower-tropospheric heating rate. According to these three types, we choose three representative experiments from section \ref{subsec:x0} to study the mechanism of the episodic deluges. Figure \ref{fig_mechanism} shows the analyses. The left column shows an example of episodic deluges with a lower-tropospheric heating rate of 0.2 K day$^{-1}$ (Figures \ref{fig_mechanism}a \& d). The middle column shows an example of episodic deluges with a negative lower-tropospheric heating rate of $-$0.2 K day$^{-1}$ (Figures \ref{fig_mechanism}b \& e). The right column shows an example of quasi-steady precipitation with a negative lower-tropospheric heating rate of $-$0.8 K day$^{-1}$ (Figures \ref{fig_mechanism}c \& f). The upper-tropospheric radiative heating rates are all $-$1.2 K day$^{-1}$, and the inhibition height is 8 km.

Saturated equivalent potential temperature ($\theta_{se}$) is conserved for a reversible moist adiabatic process, so $\theta_{se}$ is a useful measure of the static stability of saturated atmosphere \cite{emanuel1994atmospheric}. When the lower-tropospheric radiative heating rate is positive (Figure \ref{fig_mechanism}a), the $\theta_{se}$ profile has a strong inversion at $\approx$ 1 km, but has almost no obvious inversion above 2 km (red line in Figure \ref{fig_mechanism}j). The $\theta_{se}$ inversion at $\approx$ 1 km is unlikely to be important for suppressing the convection. Figures~\ref{fig_noheating}c \& d suggest that the inhibition layer reaches up to 10 km (see also Fig. 3 in \citeA{Seeley2021}). Moreover, when we remove the heating at 1--2 km in the polar night experiment, episodic deluges still exist (Figure \ref{fig_noheating}). When the lower-tropospheric radiative heating rate is negative (Figures \ref{fig_mechanism}b \& c), the $\theta_{se}$ profiles of the $-$0.2 K day$^{-1}$ case and the $-$0.8 K day$^{-1}$ case (Figure \ref{fig_mechanism}k \& l) appear broadly similar, but only the $-$0.2 K day$^{-1}$ case has episodic deluges (Figure \ref{fig_mechanism}e). Therefore, for episodic deluges, $\theta_{se}$ does not appear to yield insight into  the location of the inhibition layer. In what follows, we try to understand the underlying mechanism through a different angle.



\begin{figure}[!htbp]
\noindent
    \centering
    \includegraphics[width=1.0\textwidth]{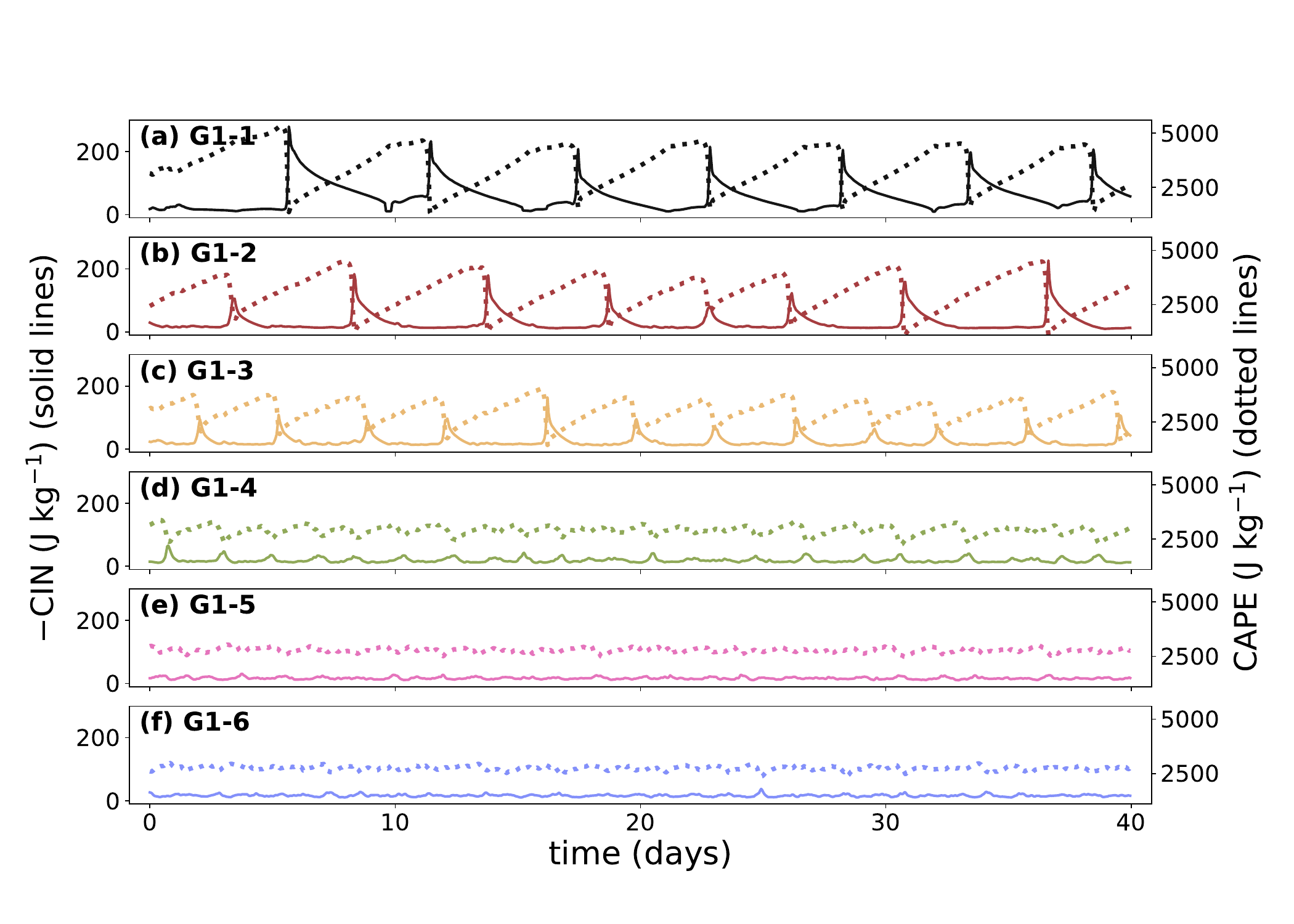}
    \caption{The time series of convective inhibition (CIN) and convective available potential energy (CAPE) for the experiments in Figure \ref{fig_lower_layers}. We plot the negative value of CIN for clarity.}
    \label{fig_cin}
\end{figure}

The difference in convective available potential energy (CAPE) and convective inhibition (CIN) at the start and end of the precipitation is large (Figures \ref{fig_mechanism}m \& p), suggesting another possible way to understand the episodic deluges. CAPE is defined as
\begin{linenomath*}
\begin{equation}
    \mathrm{CAPE} = R_{d}\int_{p(\mathrm{LFC})}^{p(\mathrm{EL})}(T_{ve}-T_{vp})d\ln(p), 
    \label{CAPE_equation}
\end{equation}
\end{linenomath*}
where $R_{d}$ is the gas constant for dry air, LFC is the level of free convection, EL is the level of neutral buoyancy, $T_{ve}$ is the virtual temperature of the environment, and $T_{vp}$ is the virtual temperature of the rising air parcel \cite{Williams_1993,Riemann_Campe_2009}. CAPE shows the total energy available for convection \cite{emanuel1994atmospheric}, but large CAPE does not guarantee the occurrence of strong convection. To release CAPE, the air parcel needs to overcome the negative buoyancy from the environment and rise high enough. CIN measures the intensity of the inhibition energy, and is defined as
\begin{linenomath*}
\begin{equation}
    \mathrm{CIN} = R_{d}\int_{p(\mathrm{SFC})}^{p(\mathrm{LFC})}(T_{ve}-T_{vp})d\ln(p), 
    \label{CIN_equation}
\end{equation}
\end{linenomath*}
where SFC is the surface. The value of CIN indicates whether convection will occur \cite{colby1984convective,Williams_1993, Riemann_Campe_2009}, and the value of CAPE indicates how strong the convection will be if convection occurs \cite{emanuel1994atmospheric,renno1996natural,Moncrieff1975}. Note that CIN is always negative. The more negative CIN is, the stronger the inhibition is.

Figure \ref{fig_cin} shows the time series of CAPE and CIN for the experiments in G1. When the precipitation pattern is episodic, CAPE decreases and CIN gets significantly more negative after the deluge. The strongly negative CIN inhibits convection. During the inhibition period, CAPE increases while CIN weakens as a function of time. When CIN is too weak to inhibit convection, convection starts, and the strong CAPE release causes an intense deluge.

\begin{table} [!ht]
 \small
 \caption{Convective inhibition (CIN) and convective available potential energy (CAPE) for all the experiments}
 \centering
 \begin{threeparttable}
 \begin{tabular}{l c l c c }
 \hline
  Group & Configuration\tnote{*}  &  CIN \ ($-2\sigma,\ +2\sigma$) & CAPE\ ($-2\sigma,\ +2\sigma$) & Precipitation\\ 
  $ $ & $ $ & J kg$^{-1}$ & J kg$^{-1}$\\
 \hline
   Exp 1 & SST = 305 K & 0 (0, 0) & 6419 (6687, 0) & quasi-steady \\
   Exp 2 & SST = 325 K & $-$40 ($-$97, $-$19) & 2991 (3819, 2054) & \textbf{episodic}  \\
   Exp 3 & polar night, SST = 330 K & $-$55 ($-$114, $-$34) & 1687 (2283, 910) & \textbf{episodic}  \\
   Exp 4 & polar night, slab ocean & $-$56 ($-$118, $-$35) & 1647 (2179, 886) & \textbf{episodic}  \\
   Exp 5 & polar night, modified heating rate & $-$61 ($-$143, $-$35) & 1649 (2589, 742) & \textbf{episodic}  \\
 \hline
   G1-1 & $\gamma_0=0.2,\ \gamma_1=-1.5,\  z_0=8$  & $-$56 ($-$177, $-$11) & 3443 (4926, 1672) & \textbf{episodic}  \\
   G1-2 & $\ \ \gamma_0=0,\ \gamma_1=-1.5,\  z_0=8$ & $-$28 ($-$115, $-$11) & 3019 (4409, 1414) & \textbf{episodic}  \\
   G1-3 & $\gamma_0=-0.2,\ \gamma_1=-1.5,\  z_0=8$  & $-$24 ($-$87, $-$12) & 2980 (3823, 1890) & \textbf{episodic}   \\
   G1-4 & $\gamma_0=-0.5,\ \gamma_1=-1.5,\  z_0=8$  & $-$19 ($-$38, $-$11) & 2899 (3257, 2392) & undetermined   \\
   G1-5 & $\gamma_0=-0.8,\ \gamma_1=-1.5,\  z_0=8$  & $-$16 ($-$25, $-$11) & 2780 (2970, 2556) & quasi-steady   \\
   G1-6 & $\gamma_0=-1.2,\ \gamma_1=-1.5,\  z_0=8$  & $-$16 ($-$25, $-$11) & 2745 (2921, 2537) & quasi-steady   \\
 \hline  
   G2-1 & $\gamma_0=-0.2,\ \gamma_1=-0.2,\  z_0=8$  & $-$8 ($-$18, $-$4) & 2811 (3015, 2507) & quasi-steady   \\
   G2-2 & $\gamma_0=-0.2,\ \gamma_1=-0.3,\  z_0=8$  & $-$15 ($-$39, $-$9) & 3213 (3593, 2549) & undetermined   \\
   G2-3 & $\gamma_0=-0.2,\ \gamma_1=-0.5,\  z_0=8$  & $-$16 ($-$63, $-$9) & 3033 (3603, 2151)) & \textbf{episodic}   \\
   G2-4 & $\gamma_0=-0.2,\ \gamma_1=-0.7,\  z_0=8$  & $-$17 ($-$54, $-$10) & 3025 (3548, 2289) & \textbf{episodic}   \\
   G2-5 & $\gamma_0=-0.2,\ \gamma_1=-1.0,\  z_0=8$  & $-$21 ($-$78, $-$11) & 2856 (3624, 1797) & \textbf{episodic}  \\
   G2-6 & $\gamma_0=-0.2,\ \gamma_1=-1.5,\  z_0=8$  & $-$23 ($-$78, $-$11) & 2940 (3756, 1912) & \textbf{episodic}  \\
 \hline  
   G3-1 & $\gamma_0=\gamma_1=-1$  & $-$17 ($-$41, $-$12) & 2750 (2916, 2554) & quasi-steady  \\
   G3-2 & $\gamma_0=\gamma_1=-2$  & $-$14 ($-$34, $-$10) & 2661 (2826, 2523) & quasi-steady  \\
   G3-3 & $\gamma_0=\gamma_1=-3$  & $-$12 ($-$39, $-$8) & 2634 (2829, 2380) & quasi-steady  \\
   G3-4 & $\gamma_0=\gamma_1=-4$  & $-$10 ($-$35, $-$7) & 2703 (3033, 2397) & quasi-steady  \\
   G3-5 & $\gamma_0=\gamma_1=-5$  & $-$9 ($-$34, $-$6) & 2746 (3142, 2324) & quasi-steady  \\
 \hline  
   G4-1 & $\gamma_0=0,\ \gamma_1=-5,\  z_0=3\  $  & $-$26 ($-$56, $-$12) & 2768 (3210, 2138) & \textbf{episodic}   \\
   G4-2 & $\gamma_0=0,\ \gamma_1=-5,\  z_0=5\  $  & $-$33 ($-$73, $-$14) & 2797 (3489, 2025) & \textbf{episodic}   \\
   G4-3 & $\gamma_0=0,\ \gamma_1=-5,\  z_0=7\  $  & $-$41 ($-$118, $-$15) & 2913 (4060, 1579) & \textbf{episodic}   \\
   G4-4 & $\gamma_0=0,\ \gamma_1=-5,\  z_0=9\  $  & $-$36 ($-$126, $-$11) & 3079 (4250, 1598) & \textbf{episodic}   \\
   G4-5 & $\gamma_0=0,\ \gamma_1=-5,\  z_0=11$  & $-$35 ($-$122, $-$13) & 3269 (4348, 1836) & \textbf{episodic}   \\
   G4-6 & $\gamma_0=0,\ \gamma_1=-5,\  z_0=13$  & $-$30 ($-$111, $-$13) & 3411 (4358, 2024) & \textbf{episodic}   \\
 \hline
 \end{tabular}
 \begin{tablenotes}
   \item[*] $\gamma_0$: the lower-tropospheric radiative heating rate in units of K day$^{-1}$. $\gamma_1$: the upper-tropospheric radiative heating rate in units of K day$^{-1}$. $z_0$: the height of the inhibition layer in units of km.  
  \end{tablenotes}
  \end{threeparttable} 
 \label{table_cin}
\end{table}

Table \ref{table_cin} shows the average and the 2-$\sigma$ control limits of CIN and CAPE for all the experiments. For the experiments with episodic deluges, lower 2-$\sigma$ is less than $-$50 J kg$^{-1}$. CAPE does not differ much between episodic or quasi-steady precipitation experiments. The experiments in group G1 show that episodic deluge cases tend to have strong CIN and CAPE. The experiments in group G2 show that large CAPE alone is not enough to trigger episodic deluges. The experiments in group G4 show that when CIN is relatively strong enough, even if CAPE is a bit small, for example, G4-1 and G4-2, episodic deluges still exist. The results are consistent with the common view that CIN indicates the strength of convective inhibition and CAPE indicates how strong the convection is \cite{colby1984convective,emanuel1994atmospheric}.

\begin{figure}[ht]
\noindent
    \centering
    \includegraphics[width=0.8\textwidth]{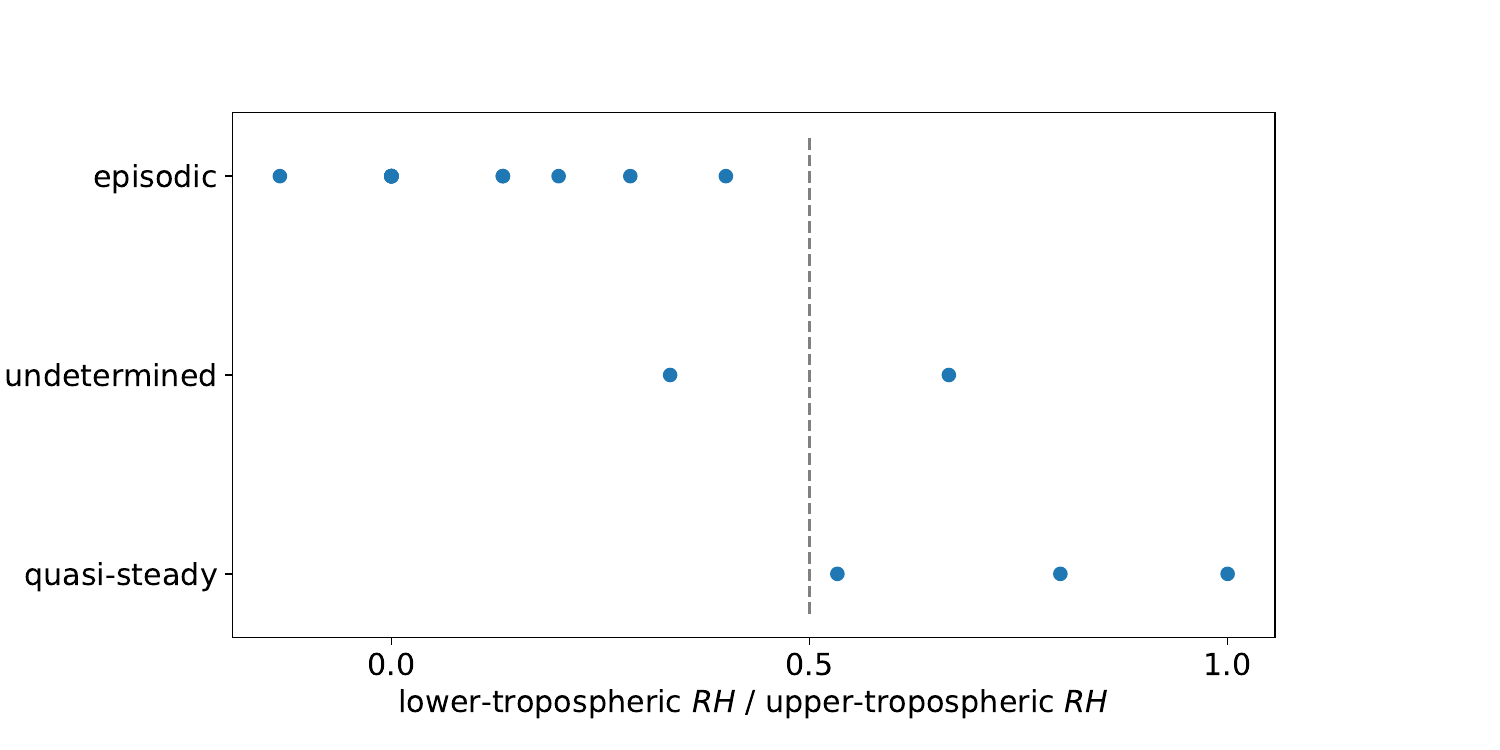}
    \caption{Relative heating rate threshold for  episodic deluges. The horizontal axis shows the ratio of the lower-tropospheric radiative heating rate to the upper-tropospheric radiative heating rate (heating rates are positive, cooling rates are negative). Each circle stands for a simulation. Precipitation is episodic when the ratio of lower-tropospheric heating rate to upper-tropospheric heating rate is less than about 0.5.}
    \label{fig_threshold}
\end{figure}

In short, the onset of episodic deluges depends on two conditions. First, the lower-tropospheric radiative heating rate should be close to zero (can be negative or positive) to maintain an inhibition period. Second, the upper-tropospheric cooling rate should be strong enough to increase the temperature lapse rate and trigger strong convection. Therefore, we use the ratio of the lower-tropospheric heating rate to the upper-tropospheric heating rate as an index of the vertical gradient in radiative cooling (Figure \ref{fig_threshold}). We find that the precipitation is episodic when this index is smaller than 0.5, and is non-episodic when this index is larger than 0.5.

Why choose the above index? Because the ratio of the lower-tropospheric heating rate to the upper-tropospheric heating rate is the key factor for creating a large vertical temperature gradient. When the lower-tropospheric heating rate is $-$0.8 K day$^{-1}$, a $-$1.5 K day$^{-1}$ upper-tropospheric heating rate is not strong enough compared with the lower-tropospheric heating rate (Figure \ref{fig_mechanism}c). The resulting temperature gradient is too small to trigger strong convection, and the energy release during convection is not large enough to start an inhibition period, so CIN and CAPE do not change much before and after precipitation (Figures \ref{fig_mechanism}o \& r). The differences in the temperature and $\theta_{se}$ profiles before and after the precipitation are also small (Figures \ref{fig_mechanism}i \& l). When the lower-tropospheric heating rate is $-$0.2 K day$^{-1}$, a $-$1.5 K day$^{-1}$ upper-tropospheric heating rate is strong enough to cause episodic deluges (Figure \ref{fig_mechanism}b). The differences in the temperature and $\theta_{se}$ profiles before and after the deluge are obvious (Figures \ref{fig_mechanism}h \& k). The big difference in CIN and CAPE before and after the deluge (Figures \ref{fig_mechanism}n \& q) is necessary to maintain the oscillation of the deluges.

\begin{figure}[ht]
\noindent
    \centering
    \includegraphics[width=1.0\textwidth]{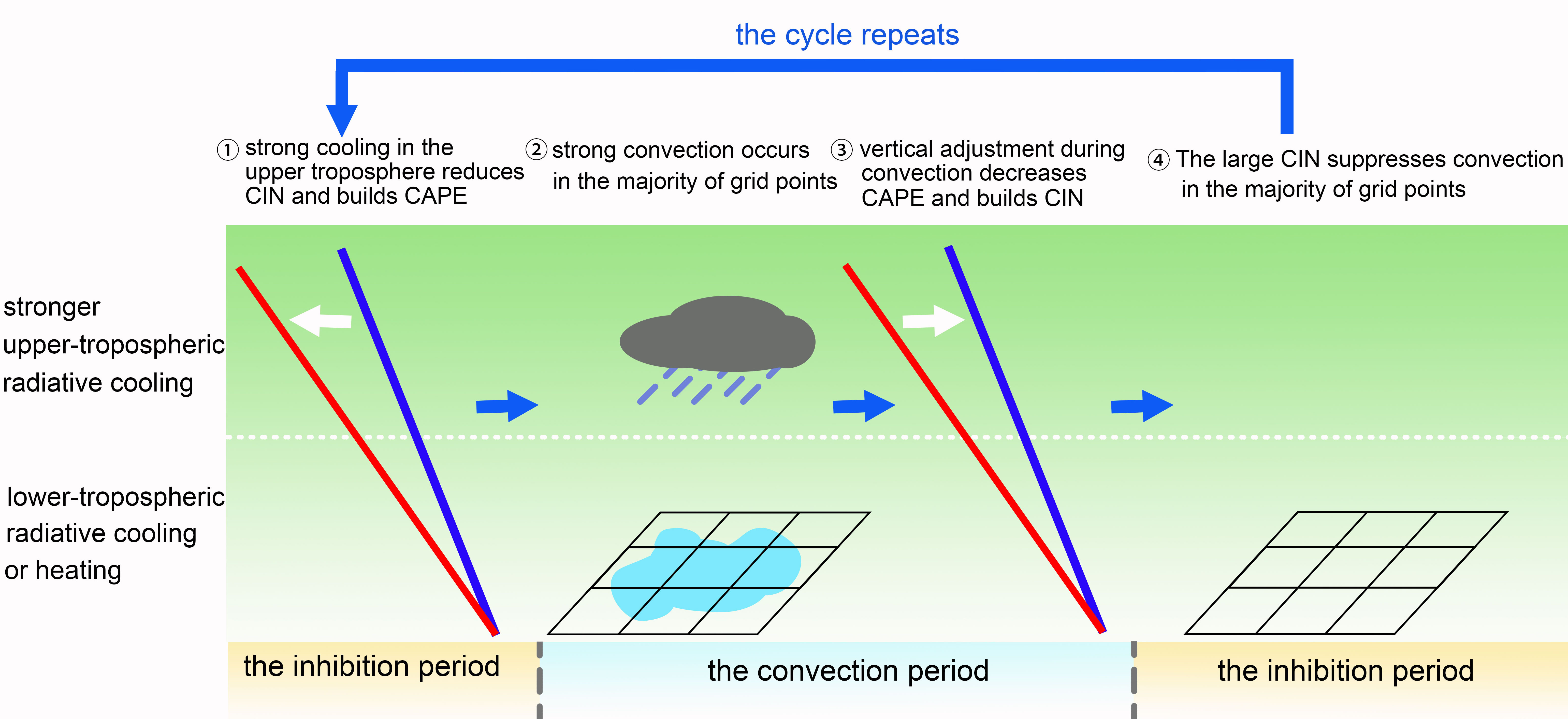}
    \caption{Schematic diagram of the processes before and after the deluge.}
    \label{fig_sketch}
\end{figure}

Figure \ref{fig_sketch} shows the schematic diagram of what happens before and after a deluge. During the inhibition period, CIN is strongly negative, suppressing convection. The strong radiative cooling in the upper troposphere ``drags" the temperature profile to a greater lapse rate, and the reevaporation in the lower troposphere also helps with weakening CIN and building CAPE (step 1). The cooling process triggers strong convection in the majority of grid points (step 2). The energy release during heavy precipitation adjusts the temperature profile to have a much smaller gradient and CIN intensifies (step 3). The majority of grid points do not precipitate under a small vertical temperature gradient and strong CIN, showing an inhibition period (step 4). Then, the cycle repeats again, with strong radiative cooling in the upper troposphere increasing the lapse rate again and triggering the next deluge.


\subsection{What determines the mean precipitation?} \label{subsec:mean_prec}
In a balanced system, the energy lost by the atmosphere should be equal to the energy gained by the atmosphere \cite{allen2002constraints,pierrehumbert2002hydrologic, o2012energetic,Xiong_2022}. In the global mean, the net effect of shortwave heating and longwave cooling on the atmosphere is balanced by the latent heat released in the atmosphere and sensible heat flux from the surface. Therefore, we can calculate the surface precipitation by

\begin{linenomath*}
\begin{equation}
    L{\rho_w}P = \frac{-c_p}{g} \int_{p_s}^{0}(HR_{LW}+HR_{SW})dp-SH, 
    \label{prec_equation}
\end{equation}
\end{linenomath*}
where $L$ is the latent heat of vaporization of water, $\rho_w$ is the density of liquid water, $P$ is the global mean surface precipitation, $c_p$ is the specific heat capacity (1004 J kg$^{-1}$ K$^{-1}$), $g$ is the gravitational constant, $p_s$ is the surface pressure, $HR_{LW}$ is the longwave heating rate in the atmosphere, $HR_{SW}$ is the shortwave heating rate in the atmosphere, and $SH$ is the sensible heat flux from the surface. $HR_{LW}$ and $HR_{SW}$ are in the units of K s$^{-1}$, and $SH$ is in the units of W m$^{-2}$. The integral starts from the surface to the top of the atmosphere.

\begin{figure}[!htb]
\centering\includegraphics[width=0.5\textwidth]{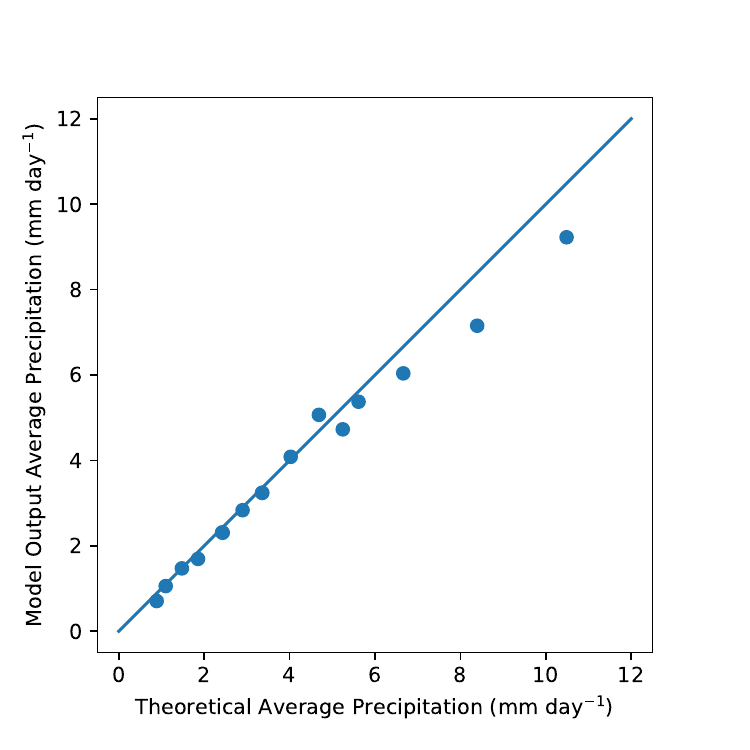}
\caption{Calculated theoretical precipitation vs actual model output (circles).}
\label{prec_cal}
\end{figure}

Figure \ref{prec_cal} shows the calculated theoretical precipitation and the actual model output. Each point stands for a single experiment. Although our experiments have fixed surface temperature, the atmosphere in most of the experiments are energy balanced. The average precipitation of the experiments lie between 1 to 12 mm day$^{-1}$, and most of the points are scattered along the $y=x$ line, showing good agreement between theoretical and simulated average precipitations. 

In the experiment with lower-tropospheric heating rate of $-$0.2 K day$^{-1}$, upper-tropospheric heating rate of $-$1.5 K day$^{-1}$, and inhibition height of 8 km (Figure~\ref{fig_lower_layers}d), the total radiative heating in the atmosphere is $-$95.4 W m$^{-2}$, the sensible heat flux from the surface is $-$1.8 W m$^{-2}$, and the total latent heating is 97.2 W m$^{-2}$. The corresponding average precipitation for this case is 3.3 mm day$^{-1}$. If the lower-tropospheric heating rate decreases to $-$0.5 K day$^{-1}$ (Figure~\ref{fig_lower_layers}e), the total radiative heating decreases to $-$115.8 W m$^{-2}$, and the average precipitation increases to 4.0 mm day$^{-1}$.

\subsection{What determines the period of the episodic deluges?} \label{subsec:period}
The simulations in sections \ref{subsec:x0}-\ref{subsec:z0} have different periods. The deluges end rather quickly, so the period of episodic deluges is essentially the time span of the inhibition period. \citeA{Seeley2021} show that precipitation reevaporation cools down the lower troposphere and breaks the inhibition. Therefore, the length of the inhibition phase should be determined by how much time the cooling process of lower troposphere takes. Let's start from the basic temperature equation \cite{vallis2019essentials}
\begin{linenomath*}
\begin{equation}    
    \frac{\partial T}{\partial t} = -(u \frac{\partial}{\partial x}+v\frac{\partial}{\partial y}+w\frac{\partial}{\partial z})T - \frac{RT}{c_v} \nabla \cdot \vec{v} + \nabla^2 (\kappa  T) + \frac{J}{c_v}, 
    \label{temp_equation}
\end{equation}
\end{linenomath*}
where $R$ is the molar gas constant (8.31 J mol$^{-1}$ K$^{-1}$); $c_v$ is the specific heat capacity (718 J kg$^{-1}$ K$^{-1}$); $\kappa$ is the thermal diffusivity representing the effect of thermal diffusion and subgrid mixing; $J$ is the external heating source including longwave cooling, shortwave heating, latent heat release through condensation, and latent cooling by reevaporation. Each experiment in this study is in a small domain and without the Coriolis force, so the horizontal temperature gradient is small, and horizontal advection ($u \frac{\partial T}{\partial x}+v\frac{\partial T}{\partial y}$) is negligible. To calculate the period, we focus on the inhibition periods, so the vertical advection ($w \frac{\partial T}{\partial z}$) and thermodynamic work ($\frac{RT}{c_v} \nabla \cdot \vec{v}$) done by the air parcel in the lower troposphere are also negligible. The influence of temperature diffusion, $\nabla^2 (\kappa T)$, is small compared with the timescale of diabatic heating ($J$). Therefore, diabatic heating alone is the main source of air temperature change. Two factors that can cool down the lower troposphere determine the period: radiative cooling and reevaporation cooling. The equation is given by

\begin{figure}[ht]
\noindent
    \centering
    \includegraphics[width=1.0\textwidth]{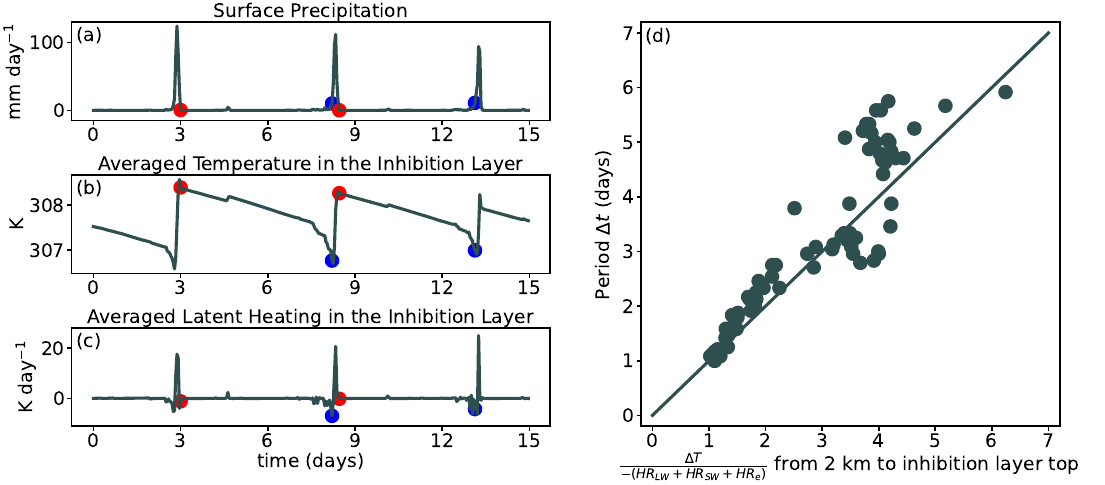}
    \caption{\textbf{Panels a--c}: Surface precipitation (a) and time series of vertically averaged temperature (b) and latent heating (c) in the inhibition layer. An inhibition period starts from the end of previous precipitation (red dot), and ends at the beginning of the next precipitation (blue dot). \textbf{Panel d}: The model output period $\Delta t$ and the calculated theoretical period (Eq. \ref{compare_equation}). Note that we remove those episodic deluge experiments with unclear boundary between inhibition periods and convection periods, such as experiments G1-4, G2-4, G4-1, and G4-2.}
    \label{fig_red.and.blue}
\end{figure}

\begin{linenomath*}
\begin{equation}
    \Delta t \approx \frac{\Delta T}{-(HR_{LW}+HR_{SW}+HR_e)},
    \label{compare_equation}
\end{equation}
\end{linenomath*}
where $-(HR_{LW}+HR_{SW})$ is the radiative cooling rate and $-HR_e$ is the reevaporation cooling rate. The inhibition period starts from the end of a deluge, and ends at the beginning of the next deluge (Figure \ref{fig_red.and.blue}a).  All the variables are the time mean and vertically averaged from 2 km to the top of the inhibition layer.


Figure~\ref{fig_red.and.blue}d compares the model output period $\Delta t$ with the calculated theoretical period (Eq. \ref{compare_equation}). Each circle stands for a single inhibition period. All circles are scattered along the $y=x$ line. This result supports the idea that radiative cooling and reevaporation cooling control the period of episodic deluges. 

The combined effect of radiative cooling and reevaporation cooling can quantitively explain the period differences in Figures \ref{fig_lower_layers}, \ref{fig_middle_layers}, and \ref{fig_inhibition_height}. In the experiments G1 with different lower-tropospheric heating rates (Figure \ref{fig_lower_layers}),  the stronger radiative cooling is, the larger the denominator in Eq. \ref{compare_equation} is, and the quicker the cooling process is. As the lower-tropospheric radiative heating rate decreases, the inhibition period shortens, until finally the inhibition period almost vanishes and the precipitation pattern becomes quasi-steady.

In the G2 group of experiments, we fix the lower-tropospheric radiative heating rate at $-$0.2 K day$^{-1}$ and change the upper-tropospheric cooling rate (Figure \ref{fig_middle_layers}). A stronger the upper-tropospheric radiative cooling results in more condensation in the upper troposphere, so more precipitation droplets reevaporate in the lower troposphere. The increasing reevaporation cooling leads to shorter period.

In the G4 group of experiments with different inhibition layer heights (Figure \ref{fig_inhibition_height}), the lower-tropospheric radiative cooling rate is zero for all the cases, so reevaporation cooling decides the periods. Higher inhibition height means weaker total radiative cooling, so less precipitation from the upper troposphere enters the inhibition layer, and therefore the period is longer and the inhibition periods are drier. 



Now we can explain the amplitude differences between the episodic deluge experiments by considering both mean precipitation and the period of episodic deluges. When controlling the mean precipitation, longer episode leads to smaller deluge amplitude. In the G1 group of experiments, when the lower-tropospheric heating rate increases, the period increases, but the mean precipitation decreases because of less total radiative cooling. The period increase outweighs the mean precipitation decrease, so the deluge amplitude increases with a larger lower-tropospheric heating rate (Figures~\ref{fig_lower_layers}b--g). In the G2 group of experiments, when the upper-tropospheric radiative cooling is stronger, the period becomes shorter, but the mean precipitation increases because of larger total radiative cooling. Both trends lead to a larger deluge amplitude (Figures~\ref{fig_middle_layers}b--g). In the G4 group of experiments, when the inhibition layer height increases while controlling other factors, the period becomes longer, but the mean precipitation decreases because of less total cooling. These two opposite trends compete with each other, so as the inhibition layer height increases, the deluge amplitude first increases then decreases (Figures~\ref{fig_inhibition_height}b--g).



\section{Summary and discussion}
We expand the episodic deluge theory developed by \citeA{Seeley2021}. We show that episodic deluges can occur during polar night, which indicates that shortwave heating is not a necessary condition for episodic deluges. Moreover, we show that episodic deluges can occur even if the lower-tropospheric radiative heating rate is negative. We perform multiple groups of experiments to show that the vertical gradient of the radiative heating rate profile is an important factor for the onset of episodic deluges. We also discuss a possible mechanism for the episodic deluges. All of our episodic deluge experiments have relatively large convective inhibition (CIN), but we cannot explain clearly how the value of CIN is linked to the inhibition mechanism. How to understand the detailed causes of the inhibition process? What exactly causes the inhibition? These questions need further studies.

The average precipitation can be understood through atmospheric column energy budget. The period and the deluge amplitude are more complex. We find three factors that influence the period of episodic deluges: lower-tropospheric radiative heating rate ($\gamma_0$), the upper-tropospheric radiative heating rate ($\gamma_1$), and the inhibition layer height ($z_0$). Generally, a higher inhibition layer, stronger radiative heating in the inhibition layer, or weaker radiative cooling in the upper troposphere leads to a shorter period. The period matches the time for radiation and reevaporation  to cool down the lower troposphere. The deluge amplitude is decided by the combined effect of the average precipitation and the period. These two trends compete with each other, so the period does not always vary monotonically with $\gamma_0$, $\gamma_1$, or $z_0$.

Another interesting factor is that the heating rate in the 325 K case are nearly horizontally uniform (Figures \ref{fig_reproduction}f \& g). The homogeneity of radiative heating rates may be an important factor for episodic deluges, as it causes a spatial coordination among the grid points, so that heavy precipitation occurs at the same time in the majority of grid points (Figure~\ref{fig_noheating}f). The coordinated dry--convection--dry cycle in the majority of grid points demonstrates an episodic deluge pattern in a macroscopic view. If the precipitation behavior among the grid points are not coordinated, convection occurs here and there at different times in different grid points, demonstrating a random, or quasi-steady, precipitation pattern in a macroscopic view. Consider, for example, the radiative heating rates in the reproduction experiments (Figures~\ref{fig_reproduction}d--g). When the surface temperature is 305 K, both lower-tropospheric heating and strong upper-tropospheric cooling exist, but not in the majority of the grid points (Figures~\ref{fig_reproduction}d \& e), so the precipitation pattern is quasi-steady. When the surface temperature is 325 K, lower-tropospheric heating and strong upper-tropospheric cooling may not be as strong as in some grid points in the 305 K case during the inhibition period, but they are more wide spread in over 65\% of the gird points (Figures~\ref{fig_reproduction}f \& g), so the precipitation pattern is episodic. These observations suggest an intimate connection between convective organization in time and space. Could episodic deluges still exist with inhomogeneous radiative heating rate? In future studies, we plan to modify the horizontal distribution of the radiative heating rate and examine how precipitation changes.

Here we study episodic deluges in a small domain, and do not include rotation and the Coriolis force, thus many weather phenomena, such as midlatitude cyclones caused by baroclinic instability, cannot be considered. In large-scale modelling, with the influence of Hadley and Walker cells, cyclones and anti-cyclones, large-scale horizontal advection is no longer negligible, and more factors, such as dynamic lifting, can possibly break  inhibition. Can episodic deluges exist in large-scale simulations?  How do episodic deluges interact with global circulation and climate? These questions are worthy of further study.

Apart from this work, two recent studies, \citeA{dagan2023convection} and \citeA{spaulding2023emergence}, also focus on the temporal variability of precipitation under hothouse climates. \citeA{dagan2023convection} find that episodic deluges cannot occur on a domain larger than $\mathcal{O}$(1000 km), and show that the propagation of gravity waves dominates the precipitation's variability in a large domain. \citeA{spaulding2023emergence} show that episodic deluges exist even in 1D single-column radiative-convective model although the period of the episodic deluges is several years, much longer than the several-day period in 3D radiative-convective model. They also show that lower-tropospheric radiative heating is not necessary for the onset of the episodic deluges. They propose a mechanism to explain the episodic deluges: ``\textit{Emergence occurs when atmospheric instability quantified by the convective available potential energy can no longer support the latent heat release of deep, entraining convective plumes.}" Whether their proposed mechanism is applicable to our 3D radiative-convective simulations here or not requires further studies.

\section{Open Research}
The cloud-resolving model SAM is publicly available at: \url{http://rossby.msrc.sunysb.edu/~marat/SAM.html}. The data in this study is publicly available at: \url{https://doi.org/10.5281/zenodo.8103889}.


\acknowledgments
We thank Marat F. Khairoutdinov for creating and maintaining SAM. We are grateful to Cheng Li, Da Yang, Daniel D. B. Koll, Feng Ding, Qiu Yang, and Zhihong Tan for the helpful discussions with them. We thank Yixiao Zhang for his help with SAM modelling. Thank Lixiang Gu, Xuelei Wang, and Quxin Cui for their help with CAPE and CIN calculation. J.Y. acknowledges support from the National Natural Science Foundation of China (NSFC) under grants 42275134, 42075046, and 42161144011.

\bibliography{agusample} 
%



%
%
%
%
%


%
%
\appendix
\section{Experiment results with discontinuous heating rate profiles}
\begin{figure}[ht]
\noindent
    \includegraphics[width=1.0\textwidth]{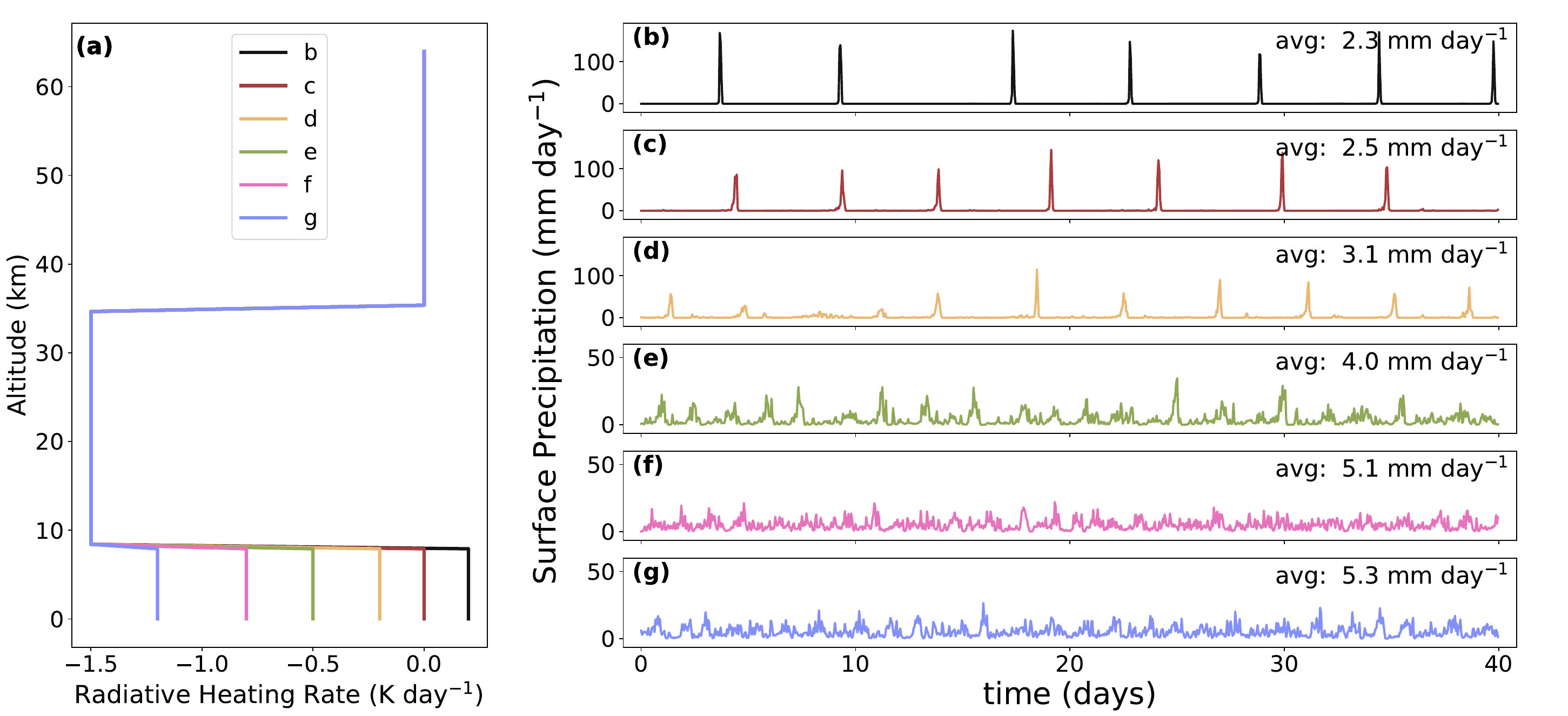}
    \caption{Same as the control group experiments of Figure \ref{fig_lower_layers}, but the heating rate is discontinuous at the inhibition layer height ($z_0$) and at the stratosphere (35 km). Panel (a) shows the prescribed radiative heating rate profiles. The upper-tropospheric heating rate is $-$1.5 K day$^{-1}$. The lower-tropospheric heating rates in panels (b) to (g) are 0.2, 0, $-$0.2, $-$0.5,  $-$0.8, and $-$1.2 K day$^{-1}$, respectively. The inhibition layer height is 8 km.}
    \label{fig_a1.x0}
\end{figure}

\begin{figure}[ht]
\noindent
    \includegraphics[width=1.0\textwidth]{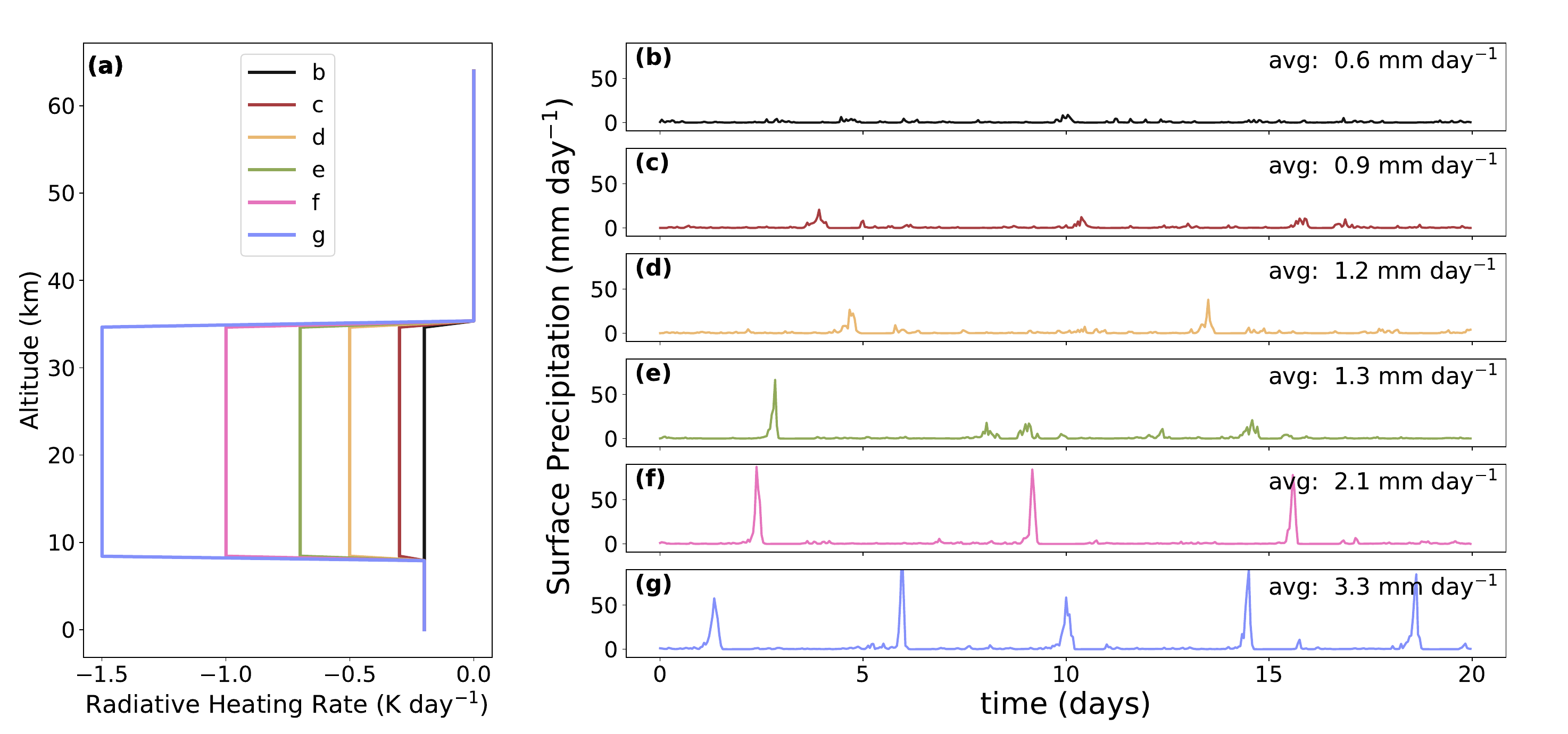}
    \caption{Same as the control group experiments of Figure \ref{fig_middle_layers}, but the heating rate is discontinuous at $z_0$ and at 35 km. Panel (a) shows the prescribed radiative heating rate profiles. The lower-tropospheric heating rate is $-$0.2 K day$^{-1}$. The upper-tropospheric heating rates in panels (b) to (g) are  $-$0.2, $-$0.3, $-$0.5, $-$0.7, $-$1.0, and $-$1.5 K day$^{-1}$, respectively. The inhibition layer height is 8 km.}
    \label{fig_a2.x1}
\end{figure}

\begin{figure}[ht]
\noindent
    \includegraphics[width=1.0\textwidth]{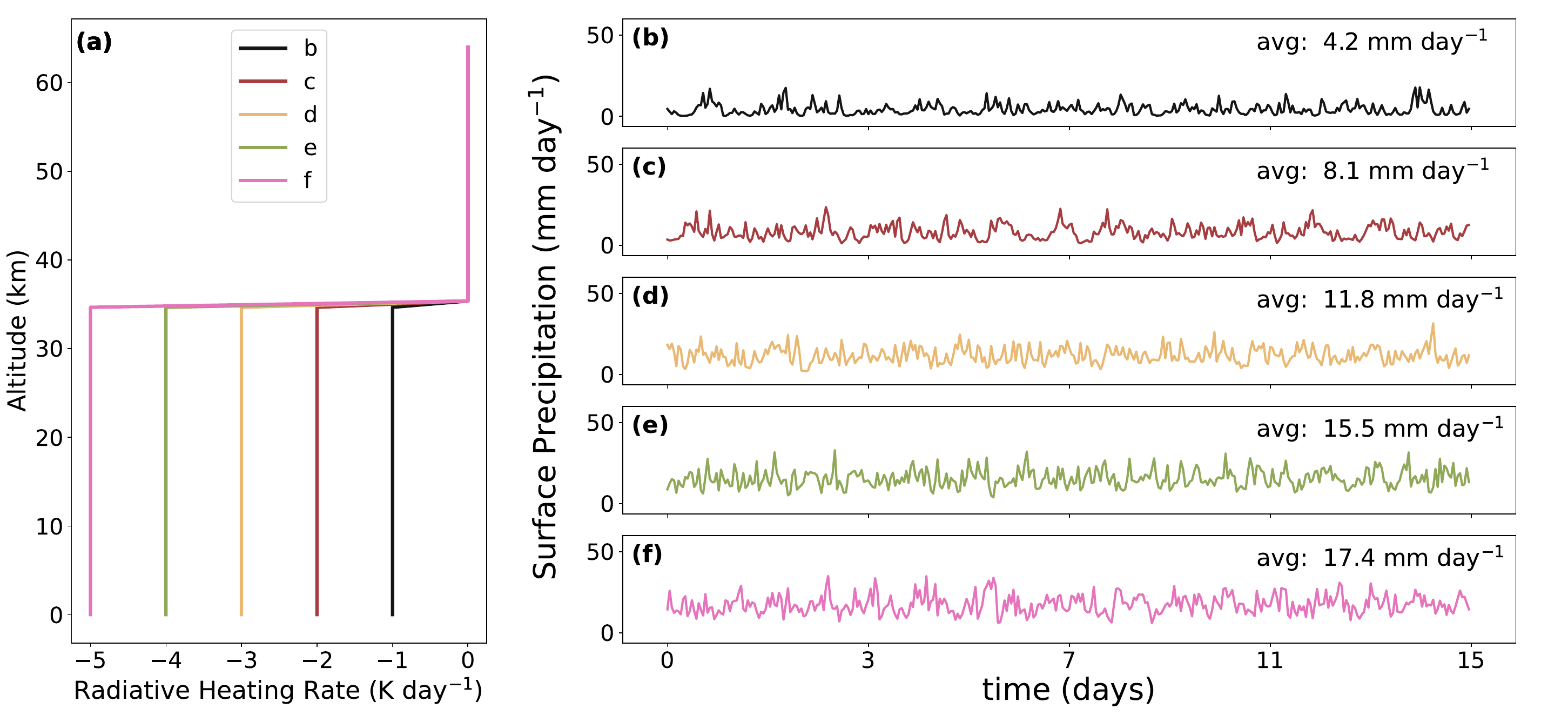}
    \caption{Same as the control group experiments of Figure \ref{fig_uni_tanh}, but the heating rate is discontinuous at $z_0$ and at 35 km. Panel (a) shows the prescribed radiative heating rate profiles. The lower-tropospheric heating rates and the upper-tropospheric heating rates in panels (b) to (f) are  $-$1, $-$2, $-$3, $-$4, and $-$5 K day$^{-1}$, respectively.}
    \label{fig_uni_stair}
\end{figure}

\begin{figure}
\noindent
    \includegraphics[width=1.0\textwidth]{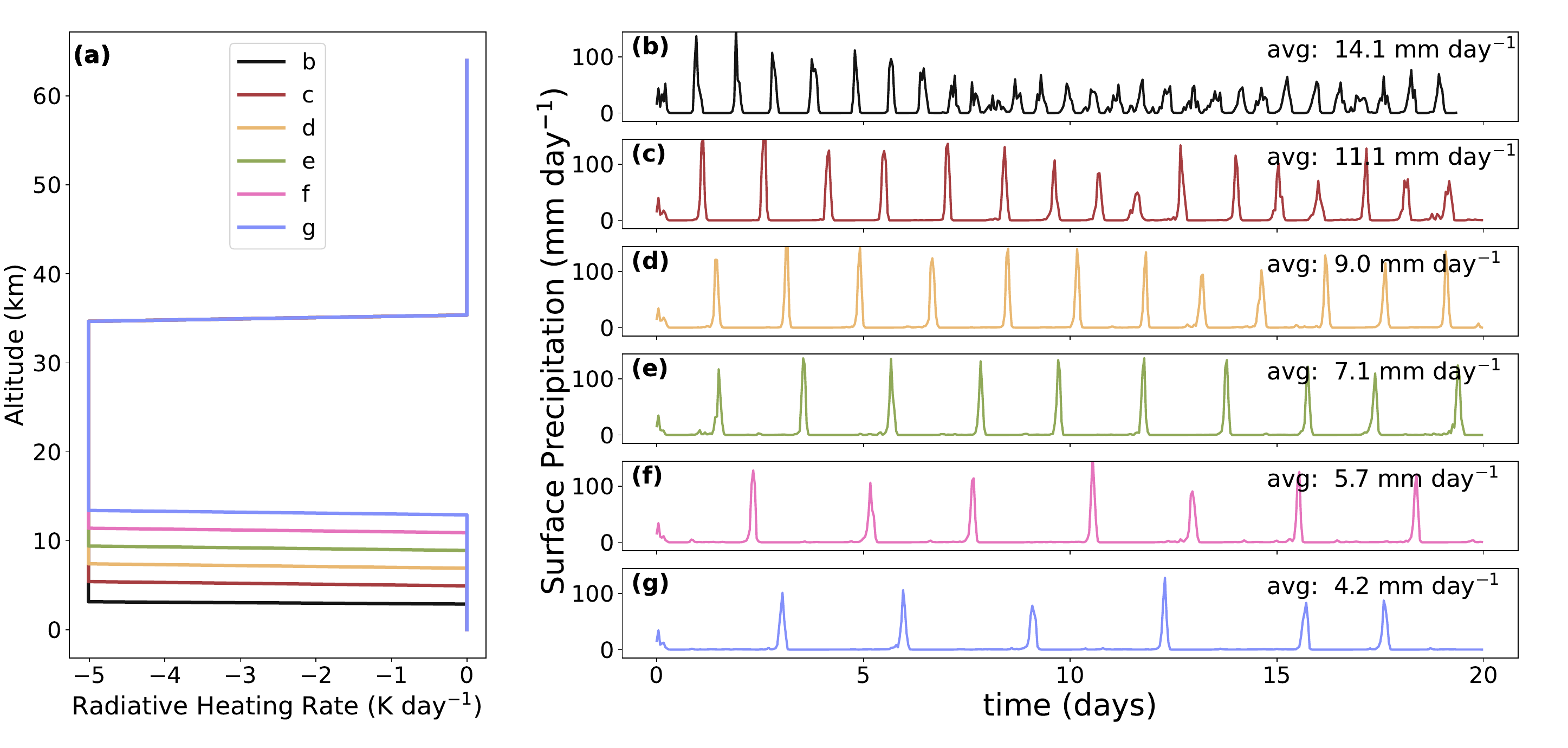}
    \caption{Same as the control group experiments of Figure \ref{fig_inhibition_height}, but the heating rate is discontinuous at $z_0$ and at 35 km. Panel (a) shows the prescribed radiative heating rate profiles. The lower-tropospheric heating rate is 0 K day$^{-1}$, and the upper-tropospheric heating rate is $-$5 K day$^{-1}$. The inhibition layer heights in panels (b) to (g) are 3, 5, 7, 9, 11, and 13 km, respectively.}
    \label{fig_a3.z0}
\end{figure}
%

\end{document}